\documentclass[aps,prb,superscriptaddress,showpacs,twocolumn,reprint]{revtex4-1}

\usepackage[pdftex]{graphicx}
\usepackage{amsmath}
\usepackage[utf8]{inputenc}
\usepackage[T1]{fontenc}
\usepackage[finnish,english]{babel}
\usepackage{verbatim}
\usepackage{float}

\begin{document}

\author{Fabian Schulz}
\affiliation{Department of Applied Physics, Aalto University School of Science, P.O.Box 15100, 00076 Aalto, Finland}
\author{Robert Drost}
\affiliation{Department of Applied Physics, Aalto University School of Science, P.O.Box 15100, 00076 Aalto, Finland}
\author{Sampsa K. H\"am\"al\"ainen}
\affiliation{Department of Applied Physics, Aalto University School of Science, P.O.Box 15100, 00076 Aalto, Finland}
\author{Thomas Demonchaux} \affiliation{Department of Applied Physics, Aalto University School of Science, P.O.Box 15100, 00076 Aalto, Finland}
\affiliation{Institut d'Electronique et de Micro\'electronique et de Nanotechnologies, IEMN, CNRS, UMR 8520, D\'epartement ISEN, 41 bd Vauban, 59046 Lille Cedex, France}
\author{Ari P. Seitsonen}
\affiliation{Institut f\"ur Chemie, University of Z\"urich, Winterthurerstrasse 190, CH-8057 Z\"urich, Switzerland}
\author{Peter Liljeroth}
\email{peter.liljeroth@aalto.fi}
\affiliation{Department of Applied Physics, Aalto University School of Science, P.O.Box 15100, 00076 Aalto, Finland}

\title{Epitaxial hexagonal boron nitride on Ir(111): A work function template}



\date{\today}

\begin{abstract}
Hexagonal boron nitride (h-BN) is a prominent member in the growing family of two-dimensional materials with potential applications ranging from being an atomically smooth support for other 2D materials to templating growth of molecular layers. We have studied the structure of monolayer h-BN grown by chemical vapour deposition on Ir(111) by low-temperature scanning tunneling microscopy (STM) and spectroscopy (STS) experiments and state-of-the-art density functional theory (DFT) calculations. The lattice-mismatch between the h-BN and Ir(111) surface results in the formation of a moir\'e superstructure with a periodicity of $\sim$29 \AA\, and a corrugation of $\sim$0.4 \AA. By measuring the field emission resonances above the h-BN layer, we find a modulation of the work function within the moir\'e unit cell of $\sim$0.5 eV. DFT simulations for a 13-on-12 h-BN/Ir(111) unit cell confirm our experimental findings and allow us to relate the change in the work function to the subtle changes in the interaction between boron and nitrogen atoms and the underlying substrate atoms within the moir\'e unit cell. Hexagonal boron nitride on Ir(111) combines weak topographic corrugation with a strong work function modulation over the moir\'e unit cell. This makes h-BN/Ir(111) a potential substrate for electronically modulated thin film and hetero-sandwich structures.

\end{abstract}

\maketitle

\section{Introduction}

Hexagonal boron nitride (h-BN) is a prominent member in the growing family of two-dimensional materials. Isostructural to graphene while being a wide bandgap insulator, h-BN has found a host of current and potential applications. These range from serving as an atomically smooth support for other 2D materials to band structure engineering in graphene/h-BN heterostructures to epitaxial growth in two-dimensional space.\cite{Dean2010,Hunt2013,Liu2014} Another area of interest in h-BN are the so-called boron nitride nanomeshes - epitaxial monolayers of h-BN grown on transition metal surfaces.\cite{Corso2004, Laskowski2007} Various studies have been motivated by their ability to act as a template for bottom-up fabrication techniques while simultaneously providing electronic decoupling from the metallic substrate.\cite{Corso2004, Corso2005, Goriachko2007, Berner2007, Dil2008, Kahle2012, Joshi2012, Schulz2013, Joshi2013} The term ``nanomesh'' highlights the peculiar structure of the h-BN monolayers: Due to the lattice-mismatch with the underlying substrate, the atomic registry between the boron and nitrogen atoms and the metallic surface is periodically modulated, resulting in a moir\'e pattern formed by regions of stronger h-BN/metal and weaker h-BN/metal interaction.

Topography, work function and chemical reactivity are periodically modulated over the moir\'e pattern, which is the origin of the templating effect of the h-BN nanomesh. Recent work on extended, self-assembled molecular layers on h-BN/Ir(111) \cite{Schulz2013} and h-BN/Cu(111) \cite{Joshi2013} has demonstrated that these templating capabilities are not only limited to structural properties but can be extended to the electronic properties of the overlayer: The work function modulation along the h-BN moir\'e unit cell causes an energy shift of the molecular resonances,\cite{Joshi2013} potentially resulting in the local charging of the molecular layer.\cite{Schulz2013} Thus, monolayers of h-BN offer a route to grow novel organic thin films or layered heterostructures, with periodically modulated electronic properties. Three criteria control the properties of the resulting films: (i) structural corrugation of the h-BN layer, (ii) work function modulation along the moir\'e unit cell and (iii) coherence length of the moir\'e pattern. A small structural corrugation facilitates the growth of defect-free overlayers and sandwich structures, while a large work function modulation allows the patterning of electronic properties. Finally, growing high-quality, large-scale layers requires the h-BN to maintain a uniform moir\'e periodicity and orientation over the entire sample size with a low density of domain boundaries.

A low structural corrugation is generally found in weakly interacting h-BN/metal systems such as h-BN/Cu(111)\cite{Preo2005, Joshi2012, Roth2013} or h-BN/Pt(111) \cite{Preo2007}. However, the weak interaction often leads to the formation of several rotational domains,\cite{Mueller2005, Morscher2006, Mueller2010, Joshi2012} as there is no preferred growth direction. For example on copper, this leads to a variation of the moir\'e periodicity from ca. 5 nm up to 14 nm \cite{Joshi2012}, depending on the rotational alignment. This problem can be by-passed by growing h-BN on strongly interacting transition metals, which typically result in single-domain growth, as well as a larger work function modulation. The prototypical, strongly interacting nanomesh systems h-BN/Rh(111) \cite{Corso2004, Berner2007, Diaz2013} and h-BN/Ru(0001) \cite{Goriachko2007, Diaz2013} offer a work function modulation of up to 0.5 eV along the moir\'e, however, accompanied by a large structural corrugation of around 1 to 1.5 \AA. A h-BN/metal system combining the advantages of strongly and weakly interacting systems would be of great value for further exploration of electronically modulated thin films.

Recent low-energy electron diffraction experiments \cite{Orlando2012} indicate the existence of a preferred orientation of h-BN on Ir(111), while previous density functional theory (DFT) calculations \cite{Laskowski2008} as well as X-ray adsorption and photoelectron spectroscopy experiments \cite{Preo2007} suggest a weaker interaction of h-BN with Ir(111) than with Rh(111) or Ru(0001). Thus, h-BN/Ir(111) constitutes a promising candidate to combine the desired properties of preferred orientation of the moir\'e superstructure, strong work function modulation and small structural corrugation. Here, we show that h-BN grown on Ir(111) indeed fulfills these criteria. Low-temperature scanning tunneling microscopy (STM) experiments demonstrate that h-BN/Ir(111) can be grown to form large domains extending across step edges, and with a preferred orientation of the h-BN layer with respect to the substrate lattice, resulting in a low spread of the moir\'e periodicity. Scanning tunneling spectroscopy (STS) indicates a modulation of the work function along the moir\'e unit cell of $\sim$0.5 eV. Our experimental findings are complemented by extensive, state-of-the-art DFT calculations. The computational results confirm the work function modulation and indicate a weak structural corrugation of the h-BN layer of 35 pm. Combining the experimental results with the simulation of the moir\'e unit cell, we can explain the work function modulation as a result of subtle changes in the registry and interaction between the h-BN and substrate atoms.

\section{Methods}
\emph{Sample preparation.} All the experiments were carried out in an ultra-high vacuum system with a base pressure of $\sim10^{-10}$ mbar. The (111)-terminated iridium single crystal was cleaned by repeated cycles of sputtering with 1.5 kV neon ions, annealing to 900 $^{\circ}$C in $5\times10^{-7}$ mbar oxygen and subsequent flashing to 1400 $^{\circ}$C. Full monolayers of h-BN were grown by thermal cracking of borazine (B$_3$N$_3$H$_6$, Chemos GmbH) at the Ir(111) \cite{Orlando2012} substrate held at a temperature of 1080 $^{\circ}$C and with a borazine pressure of $2\times10^{-8}$ mbar. With these parameters, the h-BN layer grows with a low nucleation density and forms large domains with sizes larger than the terrace width of the Ir(111) substrate.\cite{Lu2013} The resulting h-BN domains are aligned with the substrate lattice, resulting in a very uniform moir\'e periodicity of $(29.3\pm0.6)$ \AA. When growing the h-BN at higher substrate temperatures between 1100 and 1200 $^{\circ}$C, we also find misaligned domains, yielding moire periodicities down to 22 \AA.  Temperature programmed growth of the h-BN layer by preadsorption of borazine on the sample held at room temperatures and subsequent annealing results in a mosaic-growth \cite{Lu2013} of the h-BN, yielding a large spread of rotational domains, as well as defects and grain boundaries. Using sample temperatures significantly above 1200 $^{\circ}$C suppresses the h-BN growth, potentially due to an increase in desorption of nitrogen from the Ir surface and solubility of boron into the Ir bulk.
\begin{figure}[!ht]
\includegraphics [width=0.95\columnwidth]{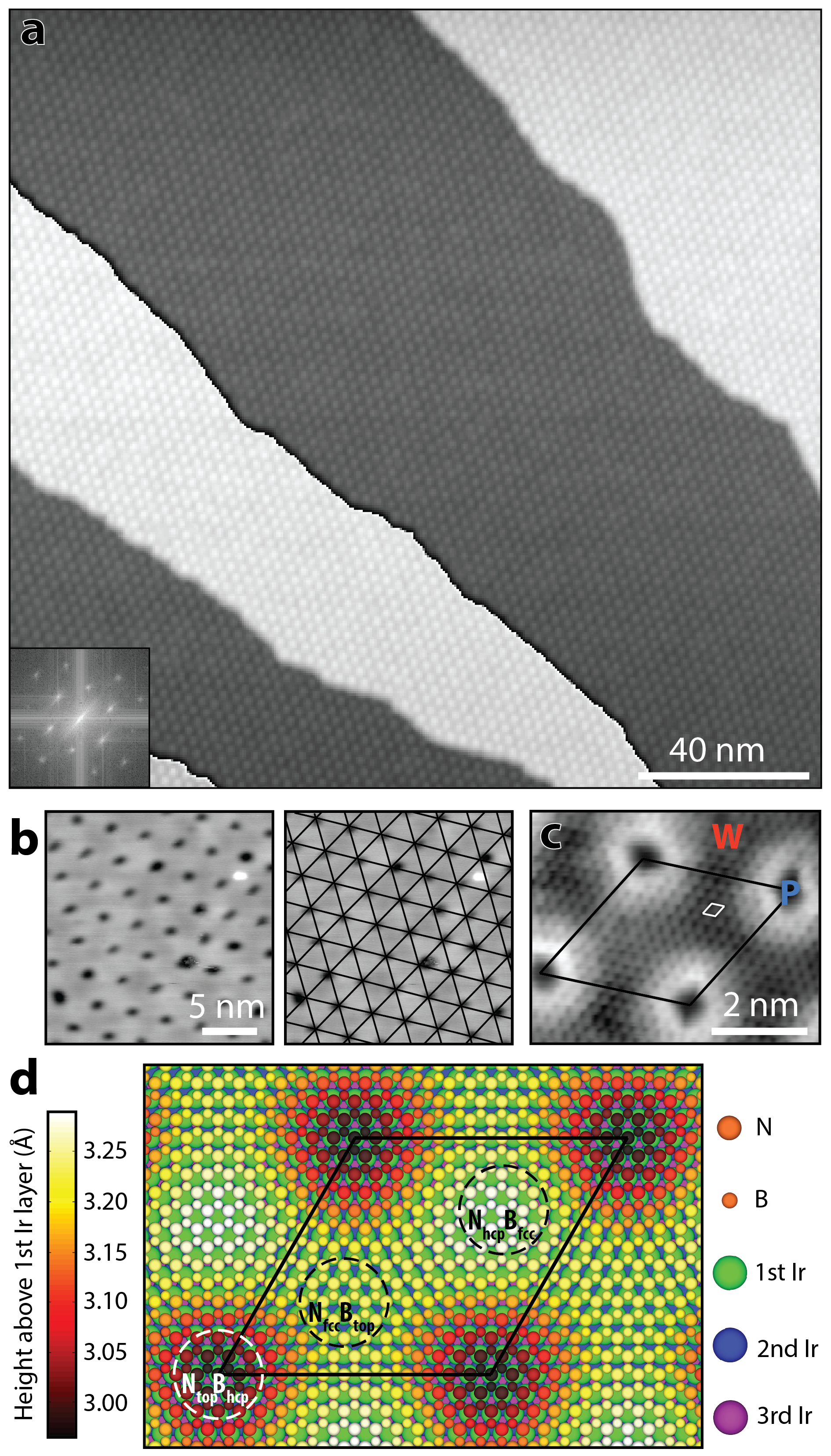}
\caption{(Color online) Moir\'e superstructure of hexagonal boron nitride on Ir(111). \textbf{(a)} STM overview image of the sample showing a single h-BN domain extending over four monatomic steps. The black-and-white color scale is adjusted to repeatedly cover the height of two terraces. The inset is a fast Fourier transform of the image, showing six sharp spots indicative of a single rotational moir\'e domain. \textbf{(b)} Zoom-in on the moir\'e pattern (left), highlighting its hexagonal periodicity with an overlayed grid (right). \textbf{(c)} Atomically resolved STM image; the moir\'e and atomic unit cells are indicated in black and white, respectively. Pore and wire regions are marked in the right panel by a blue ``P'' and a red ``W'', respectively. \textbf{(d)} DFT simulation of h-BN on Ir(111). The moir\'e unit cell as well as regions where B and N atoms occupy high-symmetry positions w.r.t. the Ir lattice are indicated. Feedback parameters: (a) 1.66 V, 0.31 nA; (b) -1.71 V, 0.40 nA; (c) 0.20 V, 3.00 nA.}
\label{fig:one}
\end{figure}

\emph{STM measurements.} After the preparation, the sample was inserted into the low-tem\-per\-a\-ture STM (Createc LT-STM) and all subsequent measurements were performed at 5 K. Differential conductance (d$I$/d$V$) spectra are recorded by standard lock-in detection on the tunneling current while sweeping the applied sample bias with a peak-to-peak modulation of 20 mV at a frequency of 517 Hz, with the current feedback loop opened at 1 V and 0.25 nA. $I(z)$ spectra were taken at a bias of 50 mV. Field emission resonances (FERs) were measured with closed feedback loop at a current setpoint of 0.5 nA. STM images and d$I$/d$V$ line spectra were processed using the \emph{WSxM} \cite{Horcas2007} and \emph{SpectraFox} \cite{SpectraFox} softwares, respectively.

\emph{DFT calculations.} The DFT calculations were performed using the QuickStep module \cite{VandeVondele2005} of the CP2K package (\verb+http://www.CP2K.org/+), where the Kohn-Sham orbitals are expanded in the basis of Gaussian functions (here, \verb+DZVP-MOLOPT-SR-GTH+ for B and N and \verb+DZVP-MOLOPT-SR-GTH-q17+ for Ir\cite{VandeVondele2007}), and plane waves up to a cut-off energy of 700~Ry, \verb+REL_CUTOFF+ of 70~Ry, for the density. Generalised Gradient Approximation (GGA) revPBE \cite{Zhang1998} was employed as the exchange-correlation functional, and the missing London dispersion incorporated using the semi-empirical DFT-D3 formalism.\cite{Grimme2010} The surface was modelled using the slab approach with four layers of the substrate of which the two at the  bottom were held fixed during the geometry relaxation, and the h-BN layer adsorbed only on one side of the slab. The length of the cell was 40~{\AA}\/ in order to decouple the two sides of the slab from each other. Only the $\Gamma$ point was used in the evaluation of the integrals over the first Brillouin zone, and the bulk lattice constant of 3.801~{\AA} obtained with revPBE-D3 was used. The occupation numbers are broadened using the Fermi-Dirac distribution at 300~K around the Fermi energy. The STM images were simulated using the Tersoff-Hamann model \cite{Tersoff1985} with an s wave tip. Further details on the method can be found in Ref. \onlinecite{Diaz2013}

\section{Results and Discussion}

Figure \ref{fig:one}a shows a large-scale STM image of monolayer h-BN grown on the Ir(111) surface by chemical vapor deposition (CVD) using borazine as the precursor (see Methods for details). A single domain extends over several monatomic steps, indicating the high quality of the h-BN layer as it extends over step edges in a carpet-like fashion.\cite{Coraux2008, Sutter2008, Lu2013} The moir\'e superstructure due to the lattice-mismatch (Ir(111): $a$ = 2.714 \AA \,\cite{Singh1968} and h-BN: $a$ = 2.505 \AA\, \cite{Pease1952}) is highlighted in Figure \ref{fig:one}b; it is formed by depressions arranged in a hexagonal lattice. The periodicity of the superstructure is indicated in the right panel of Figure \ref{fig:one}b by a hexagonal grid overlayed onto the STM image. Throughout this article, we will use the common terminology and refer to the depressions of the moir\'e as \emph{pores} and to the surrounding regions as \emph{wires}. However, it is important to note that the h-BN forms a continuous, closed monolayer without any voids. The moir\'e unit cell as well as the atomic unit cell of the h-BN are depicted in the high resolution STM topograph in Figure \ref{fig:one}c. As can be seen, the two unit cells are aligned, without any appreciable rotation with respect to each other. Since the angle between the moir\'e and h-BN lattice vectors represents a roughly ten-fold magnification of the rotation between the lattice of h-BN and Ir(111),\cite{NDiaye2008} we conclude that the misalignment between the two atomic lattices is less than $\pm$0.5 $^{\circ}$. When growing the h-BN layer at a substrate temperature of 1080 $^{\circ}$C, we only find aligned h-BN with a moir\'e periodicity of $(29.3\pm0.6)$ \AA. The small deviation from the theoretical moir\'e periodicity of 32.5 \AA\, suggests that the h-BN lattice is strained by approximately 0.8 \%, similar to the case of graphene on Ir(111).\cite{Blanc2012, Hattab2012}

A notable feature in the STM image is a bright rim around the pores. This rim appears only at a certain sample bias range and is a direct consequence of the different atomic registry within the moir\'e unit cell leading to a modulation of the electronic properties of the h-BN overlayer, as will be shown later. On the basis of such high-resolution images and previous results by Orlando \emph{et al.},\cite{Orlando2012} we have performed a dispersion-corrected DFT simulation for a 13$\times$13 on 12$\times$12 h-BN/Ir(111) unit cell, \emph{i.e.} along the vector of the moir\'e unit cell 13 h-BN unit cells occupy 12 substrate unit cells (see Methods for details). The fully relaxed theoretical h-BN/Ir(111) structure shown in Figure \ref{fig:one}d reproduces the pore and wire pattern experimentally observed in our STM experiments. From the simulation, we find the depressions of the moir\'e corresponding to a registry with the center of the B-N hexagon over a \emph{fcc} hollow site of the Ir(111) lattice, the nitrogen atom sitting on a top site and the boron atom on a \emph{hcp} hollow site (B$_{hcp}$N$_{top}$). The minimum distance between the topmost iridium layer and the h-BN lattice in this configuration is 2.95 \AA. The maximum distance between the iridium and the h-BN is found on the wire, when N atoms occupy \emph{hcp} hollow sites and B atoms \emph{fcc} hollow sites (B$_{fcc}$N$_{hcp}$), the center of the hexagon thus being on top the underlying Ir atoms. At this registry, the h-BN-Ir distance is 3.30 \AA, giving a total corrugation within the moir\'e superstructure of 35 pm. When the nitrogen sits on a \emph{fcc} hollow sites and the boron on top sites (B$_{top}$N$_{fcc}$, the center of the hexagon thus on a \emph{hcp} hollow site), the distance to the iridium layer is slightly smaller, \emph{i.e.} 3.20 \AA.  On the rim of the pore at the transition towards the wire, boron atoms occupy predominantly bridge positions and nitrogens off-center top positions (B$_{bri}$N$_{top}$). While the correspondence between the atomic registry and the different areas of the moir\'e unit cell is similar to previous findings for other h-BN/metal systems,\cite{Diaz2013} the corrugation is much smaller than in the strongly interacting systems such as Rh(111) \cite{Corso2004, Berner2007} or Ru(0001) \cite{Goriachko2007}, where the difference between pore and wire regions is around 1 \AA.\cite{Diaz2013} With 35 pm, the corrugation of the h-BN/Ir(111) system is comparable to the one found for graphene on Ir(111) (ca. 45 pm).\cite{Sampsa2013} In addition, the calculated minimum height of the h-BN monolayer above the Ir surface is close to the value of the interlayer distance in bulk h-BN of 3.33 \AA,\cite{Pease1952} while for h-BN/Rh(111) and h-BN/Ru(0001) the minimum height is  between 2 and 2.5 \AA.\cite{Diaz2013} This is clear indication that the interaction between h-BN and the Ir(111) surface is much weaker compared to the prototypical nanomesh systems. However, these subtle differences in atomic registry along the moire\'e unit cell give rise to noticeable variations of the electronic properties along the h-BN layer.
\begin{figure}[!htbp]
\includegraphics [width=.95\columnwidth]{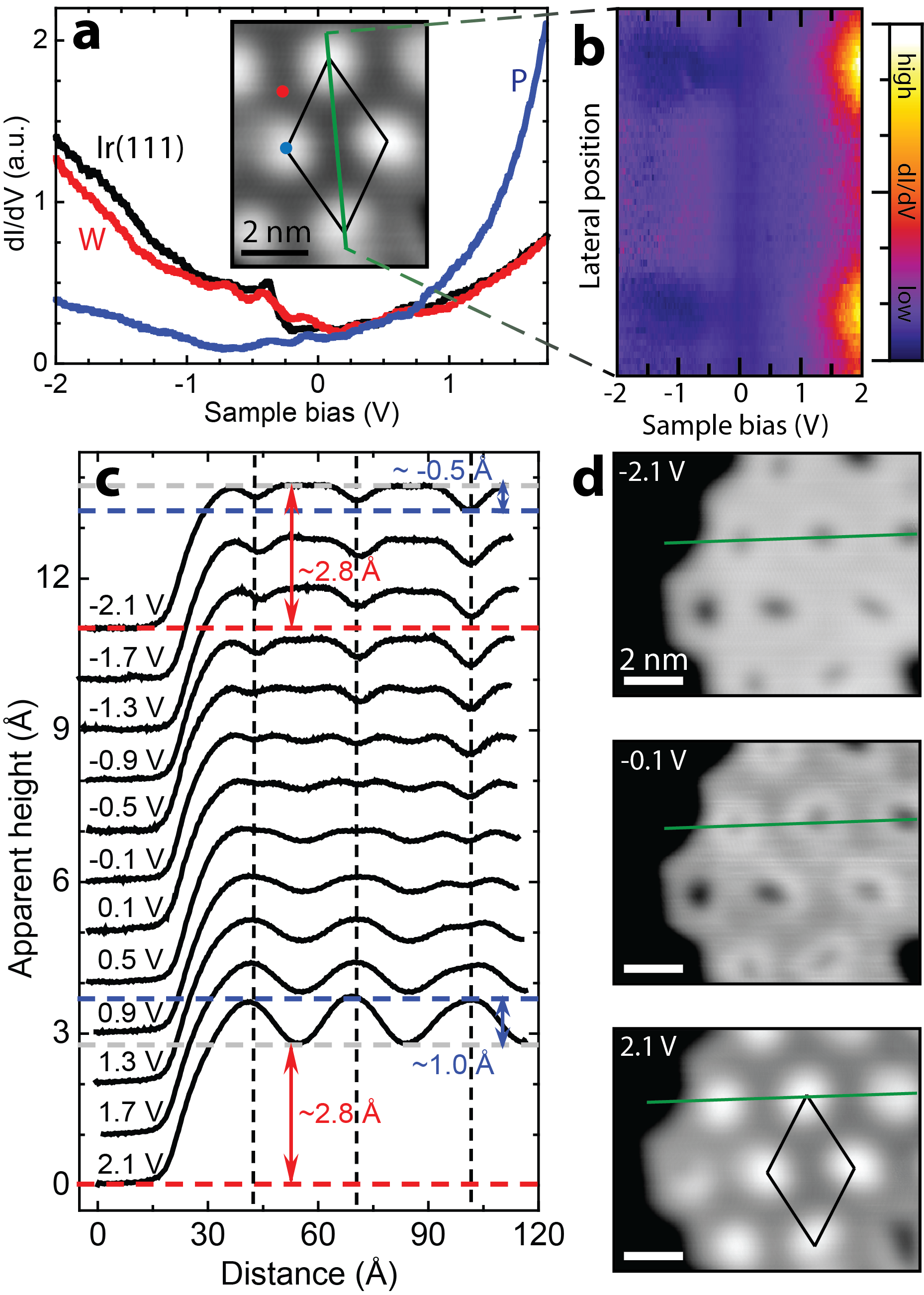}
\caption{(Color online) Experimental electronic structure of h-BN on Ir(111). \textbf{(a)} d$I$/d$V$ spectra taken on the pore and on wire of the moir\'e. \textbf{(b)} Color-coded  d$I$/d$V$ spectra taken along a line connecting two pores. \textbf{(c)} and \textbf{(d)} Bias-dependent STM contrast of the h-BN layer. Feedback parameters: inset (a) 2.10 V, 0.25 nA; (d) 0.25 nA for all images, sample bias as indicated.}
\label{fig:two}
\end{figure}

Figure \ref{fig:two}a compares differential conductance (d$I$/d$V$) spectra measured on the bare Ir(111) and on the pore and wire regions of the h-BN moir\'e. The reference spectrum of the bare iridium shows a step-like increase in the conductivity at $\sim$-380 mV due to the hole-like surface state present at the (111)-terminated Ir surface.\cite{Varykhalov2012, Altenburg2012} Metallic surface states are well-known to be very sensitive to any kind of adsorbates. Depending on the nature and the strength of the interaction between surface and adsorbate, the binding energy of the surface state can shift,\cite{Louie1978, Souk1985, Park2000, Hoevel2001, Forster2003} its onset can broaden,\cite{Park2000} and its intensity can be attenuated\cite{Jonker1981, Souk1985, Forster2003, Nicoara2006} or eventually be completely quenched.\cite{Eberhardt1983, Tzeng2000, Hoevel2001, Torrente2008} The d$I$/d$V$ spectrum taken on the wire region of the h-BN moir\'e exhibits a step-like feature around $\sim$-360 mV, which we assign to the surface state. The fact that the surface state survives underneath the wire region and does not shift significantly, confirms the weak interaction of this part of the moir\'e unit cell with the Ir(111) surface. We note, however, that the onset is broadened, suggesting a decreased lifetime of holes in the surface state.\cite{Kliewer2000} In contrast, the spectrum taken on the pore does not show any features related to the surface state. It thus appears to be quenched (or at least strongly shifted) by the stronger interaction of the h-BN layer with the metallic surface on the pore region of the moir\'e unit cell. Instead, we find a sharp rise in the d$I$/d$V$ signal at positive bias, starting around 1 V. The evolution in differential conductance along the moir\'e unit cell is depicted in Figure \ref{fig:two}b, where a series of individual d$I$/d$V$ spectra is shown as a function of lateral position in a two-dimensional colour plot. The strong increase at positive bias is observed exclusively at the pore, thus its origin must lie in the interaction of the h-BN with the metal substrate.
\begin{figure}[!hbp]
\includegraphics [width=.95\columnwidth]{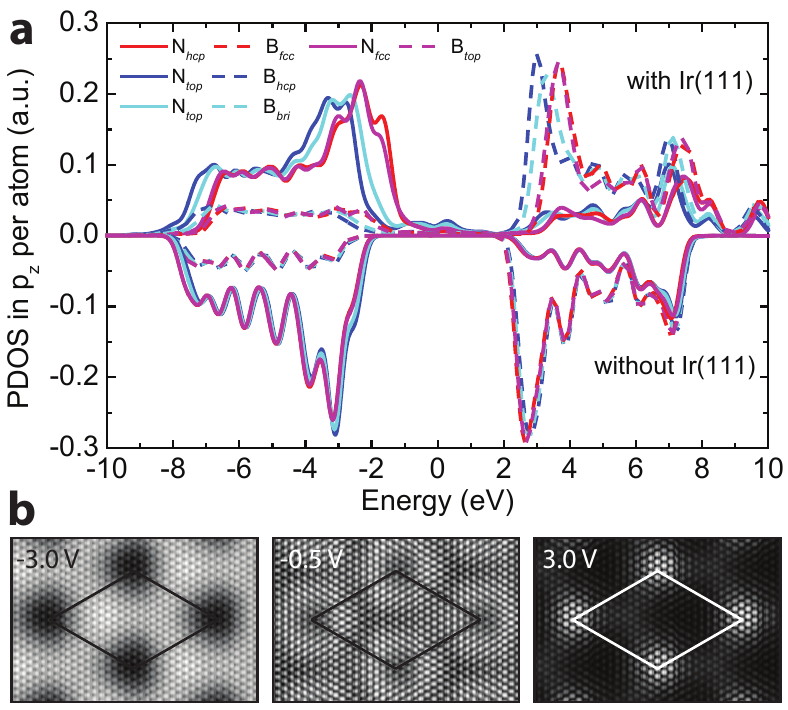}
\caption{(Color online) Theoretical electronic structure of the h-BN monolayer on Ir(111). \textbf{(a)} The projected density of states of the $p_{z}$ orbitals of boron and nitrogen resulting from a DFT calculation. \textbf{(b)} Simulated STM images at different bias voltages.}
\label{fig:three}
\end{figure}

The different features of the d$I$/d$V$ spectra are also reflected in the STM image contrast. Figure \ref{fig:two}c shows STM line profiles of the h-BN moir\'e extracted from images taken at different sample biases showing clean Ir(111) as well as an h-BN island. The series goes from -2.1 V to 2.1 V and the profile crosses two moir\'e unit cells, as indicated by the green lines in the STM images in Figure \ref{fig:two}d. At negative bias, the pores appear as depressions with an apparent depth of 0.3 - 0.5 \AA, in agreement with the actual topographic corrugation obtained from the DFT simulation (Figure \ref{fig:one}d). However, at large positive bias the moir\'e contrast inverts and the pores appear as protrusions with an apparent height of up to $\sim$1 \AA\, with respect to the wire region. This can be directly related to the increased DOS on the pore region at high bias as depicted in the spectra in Figure \ref{fig:two}a and b. At low biases, a bright rim appears around the pores with an apparent height larger than the wire or the pore. Note that the apparent height of the wire region with respect to the clean Ir(111) does not show any significant change over the entire bias region, being around $\sim$2.8 \AA. This is in rough agreement with the h-BN-Ir(111) distance on the wire given by DFT of 3.3 \AA\, (Figure \ref{fig:one}d).

To further elucidate the origin of this contrast reversal, we have plotted in Figure \ref{fig:three}a the projected density of the states (PDOS) of the $p_{z}$ orbital for boron and nitrogen atoms in the h-BN layer with ('positive' PDOS) and without ('negative' PDOS) the Ir(111) surface as obtained from our DFT calculations. The PDOS is split into different atomic registries with respect to the iridium surface, corresponding to different areas of the moir\'e unit cell. As expected, irrespective of the presence of the surface, the occupied states are dominated by the nitrogen atoms (full lines) and the unoccupied ones by the boron atoms (dashed lines). At the pore and at its rim, where the registries are B$_{hcp}$N$_{top}$  and B$_{bri}$N$_{top}$, the entire PDOS is shifted to lower energies, \emph{i.e.} the onset of the conduction band is observed at lower energies than at the wire region, causing the pores to appear bright at positive bias. As the valence band is shifted downwards as well, it becomes accessible only at larger negative bias compared to the wire, resulting in the pores being imaged as depressions and thus reversing the moir\'e STM contrast. Apart from the downward shift of the bands on the pore region, there is finite PDOS around the Fermi energy for the h-BN in Ir(111), indicating a partial - albeit small - hybridization of h-BN $p_{z}$ states with electronic states of the underlying metal substrate (most likely with the partially filled $d$ states of the iridium,\cite{Preo2007, Laskowski2008, Orlando2012}) in agreement with previous experimental findings.\cite{Preo2007} Again, this effect is the strongest for the registries corresponding to the pore and rim regions of the moir\'e, in accordance with the increased interaction with the substrate. At small energies around zero, the rim and the pore show the largest DOS, which in conjunction with the actual topographic corrugation causes the rim to appear the brightest at small biases. We can reproduce the observed contrast changes by using our DFT results to simulate STM images, following the Tersoff-Hamann model.\cite{Tersoff1985} Figure \ref{fig:three}b depicts a set of such simulations, for sample biases of $-3.0$, $-0.5$ and $3.0$ V. The images at large negative and large positive bias yield excellent agreement with the experimental observations, as the appearance of the pore switches from a depression to a protrusion. Also the bright rim around the pore at low voltages is well reproduced by our STM simulation.
\begin{figure}[!tbp]
\includegraphics [width=.9\columnwidth]{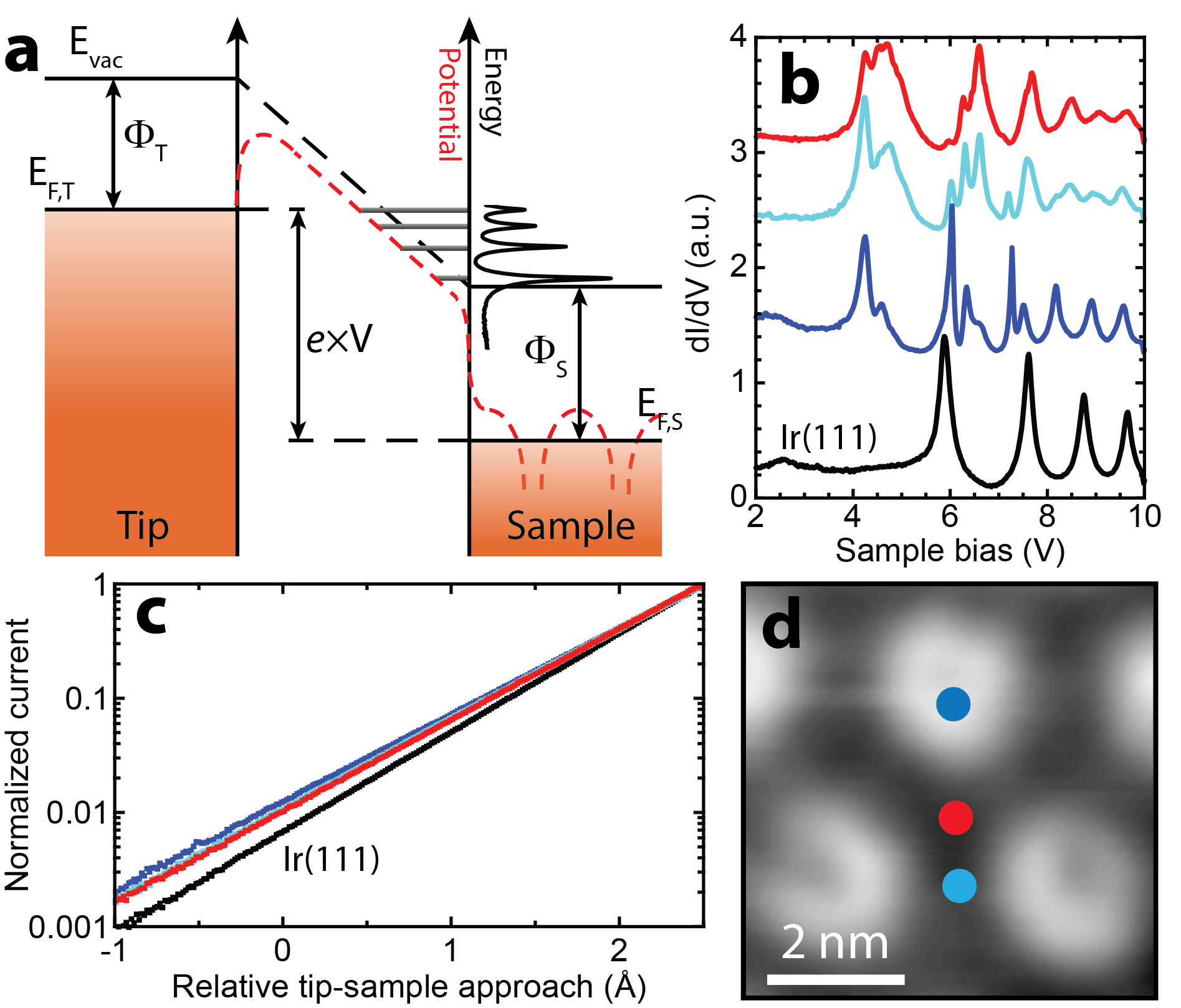}
\caption{(Color online) Field emission resonances and $I(z)$ spectroscopy on h-BN/Ir(111). \textbf{(a)} STM junction under high bias in the FER regime. \textbf{(b)} FER and \textbf{(c)} $I(z)$ point spectroscopy measured on the clean Ir(111) (black) and on different parts of the h-BN moir\'e. FER spectra vertically offset for clarity. \textbf{(d)} STM topography indicating the location of the point spectra in panels b and c. Feedback parameters: (d) 0.20 V, 1.00 nA.}
\label{fig:four}
\end{figure}

These variations in the electronic structure of the h-BN layer suggest a possible modification of the local tunneling barrier along the moir\'e unit cell. The tunneling barrier is directly related to the local work function or surface potential, which can be probed with high spatial resolution by measuring field emission resonances in the STM junction.\cite{Binnig1985, Becker1985} FERs, also known as Gundlach oscillations \cite{Gundlach1966} arise in the regime of Fowler-Nordheim tunneling,\cite{Fowler1928} \emph{i.e.} when the applied bias is larger than the sample work function and thus, the tip Fermi level is above the vacuum level of the sample. The trapezoidal potential due to the drop of the bias voltage along the tunneling junction can give rise to hydrogen-like electronic resonances confined in the vacuum junction by the sample surface and the classical turning point at that trapezoidal potential, as depicted schematically in Figure \ref{fig:four}a. Qualitatively, these resonances can be understood as image potential states under an external electric field. As the energy of these resonances depends on the local work function of the sample \cite{Koles2000}, FERs have found wide application in the STM community to map work function changes, in particular of thin films grown on metal substrates such as oxide films \cite{Rienks2005, Koenig2009}, thin insulating layers of NaCl \cite{Pivetta2005, Ploigt2007} and CuN \cite{Ruggiero2007} or monolayers of graphene \cite{Wang2010} and h-BN \cite{Joshi2012}. To a first approximation, the sample work function is given by the energy of the first FER. Figure \ref{fig:four}b shows FER spectra measured at different parts of the h-BN moir\'e as well as on the clean Ir(111) for comparison.

On the bare iridium, the first FER appears at $\sim$5.8 V, in good agreement with the work function of Ir(111) of 5.76 eV.\cite{Michaelson1977} In contrast, the h-BN yields its first FER at $\sim$4.2 to 4.4 V, indicating a strong reduction of the work function on the h-BN overlayer of around 1.4 eV. This overall work function reduction is comparable with previously reported values for other h-BN/metal systems.\cite{Morscher2006, Goriachko2007, Joshi2012} Interestingly, the first three FERs of the h-BN layer show some internal structure, being actually composed of three subpeaks, whose relative intensities vary depending on the area within the moir\'e unit cell. Such an effect has been observed previously when measuring FERs over the moir\'e of one monolayer of FeO on Pt(111) and was attributed to contributions from different parts of the moir\'e unit cell.\cite{Rienks2005} It has been pointed out that since FERs are measured in a closed feedback configuration and at large bias (thus at a large tip-sample distance) the effective area that is probed can be in the order of 100 \AA$^2$.\cite{Becker1985, Rienks2005} The subpeaks are most clearly resolved in the second FER of the h-BN. In fact, it has been shown that the energetic position of the second FER is a good measure to determine relative shifts of the surface potential on the local scale, as it is less influenced by the image potential at the sample surface.\cite{Lin2007} Inspecting the internal structure of the second FER for the different parts of the h-BN moir\'e, we find the three peaks being located at $\sim$6.0, 6.3 and 6.6 V. The subpeak of lowest energy has its maximum intensity at the pore, while the peak of highest energy shows maximum intensity on the wire region, indicating a reduction of the work function when going from the wire to the pore. This result is supported by measurements of the local barrier heights using $I$($z$) spectroscopy. The tunneling current decreases exponentially with increasing tip-sample distance, whereby the decay constant is proportional to the square root of the apparent barrier height. Normalized $I(z)$ spectra, taken on clean Ir(111) and on different parts of the h-BN moir\'e, are plotted in Figure \ref{fig:four}c; it clearly can be seen that on both the pore and wire, the current decays much more slowly than on the clean iridium, indicating a lower apparent tunneling barrier. Approximating the barrier height $\Phi_{b}$ as the average work function of the tip-sample system, \emph{i.e.} $\Phi_{b} = (\Phi_{t} + \Phi_{s})/2$,\cite{Kern2010} we can deduce from an exponential fit and via $\Delta\Phi_{s} = 2(\Phi_{Ir} - \Phi_{hBN})$\cite{Kern2010} an overall work function reduction w.r.t. the iridium surface of $\sim$1.6 eV for the wire and $\sim$1.2 eV for the pore (note that $\Phi_{Ir}$ and $\Phi_{hBN}$ refer to the potential barriers as determined from the exponential fit). Combining the results of the FER and $I(z)$ measurements, we find a modulation of the work function within the moir\'e of roughly 0.5 eV.

\begin{figure}[htbp]
\includegraphics [width=.9\columnwidth]{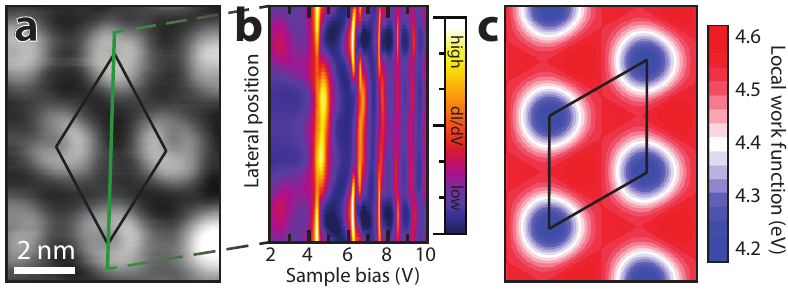}
\caption{(Color online) Mapping the work function changes over the moir\'e unit cell. \textbf{(a)} STM topography image showing the location of the FER line spectra plotted in panel b. \textbf{(b)} Color-coded d$I$/d$V$ FER spectra taken along the line indicated in panel a. \textbf{(c)} Calculated work function changes over the moir\'e unit cell. Feedback parameters: (a) 0.20 V, 1.00 nA.}
\label{fig:five}
\end{figure}

For a more detailed mapping of the work function variations, we measured FER spectra along the high-symmetry line of the moir\'e unit cell connecting two next-nearest neighbor pores with each other (Figure \ref{fig:five}a). The result is displayed as color-coded d$I$/d$V$ intensity as a function of lateral position and applied sample bias in Figure \ref{fig:five}b. At the pores, the second FER yields its highest intensity at $\sim$6.2 V, with little variation along the area of the pore. On the wire region, the maximum intensity is found between $\sim$6.6 and 6.7 V, with a slight asymmetry between the upper and lower part of the wire. Interestingly, the transition from low work function at the pore to high work function at the wire appears rather sudden, indicated by the co-existence region of the two corresponding peaks instead of a smooth shift of the second FER towards higher energy.

To compare the experimentally measured variations in the local surface potential with our DFT calculations, we have plotted the calculated Hartree potential above the h-BN layer in Figure \ref{fig:five}c, which is approximately equal to the local work function. The simulation confirms that the pores of the h-BN moir\'e yield the lowest work function of $\sim$4.2 eV; the highest work function of $\sim$4.6 eV is found at the wire region that corresponds to the B$_{fcc}$N$_{hcp}$ registry, while the B$_{top}$N$_{fcc}$ wire region has a slightly lower work function. Thus, the variation of the work function follows the modulation in the interaction strength between the different regions of the h-BN moir\'e and the iridium substrate, allowing us to relate the observed changes to the calculated PDOS of the B and N $p_{z}$ orbitals. First, we note that the overall reduction of the work function - confirmed by our DFT calculations to be more than 1 eV - can be explained by the `push-back' or `pillow' effect:\cite{Ishii1999, Hwang2009} Due to the large spill-out of the electronic wave functions at a metal surface, an interface dipole-layer pointing into the vacuum is formed.\cite{Smoluchowski1941} This dipole-layer is known to have a significant contribution to the work function of metals (as it points towards the vacuum, it increases the magnitude of the work function).\cite{Smoluchowski1941, Lang1971, Nieminen1976} However, upon formation of the h-BN layer, the wave function spill-out is strongly reduced (`push-back') due to the Pauli exclusion principle, resulting in a reduction of the work function. Now, to explain the variations of the work function within the h-BN moir\'e unit cell, we recall that the PDOS plotted in Figure \ref{fig:three}a indicates a hybridization of $p_{z}$ orbitals of the nitrogen atoms with states of the underlying metal substrate. This implies a re-distribution of electron density from the h-BN layer across the interface towards the metal substrate, \emph{i.e.} the h-BN layer becomes slightly positively charged. As a result, an interface dipole pointing towards the metal substrate is formed and reduces the work function further. This hybridization effect is the strongest at the pore where N occupies top positions and the interaction between h-BN and Ir(111) is maximum, thus the work function is the lowest on the pore. On the wire, in particular on the least interacting regions where the registry is B$_{fcc}$N$_{hcp}$ and B$_{top}$N$_{fcc}$, the hybridization is minimal and thus, this is the region with the largest work function within the moir\'e unit cell. Overall, the calculated work function yields very good agreement - qualitatively and quantitatively - with our FER measurements.

\section{Conclusions}
In summary, we have provided a detailed description of the structure of monolayer h-BN grown on Ir(111) at the atomic level. Due to the lattice mismatch between the h-BN and the Ir(111), a moir\'e superstructure with a periodicity of $\sim$29 \AA\, and a corrugation of $\sim$0.4 \AA\, is formed. The strongly interacting pores of the moir\'e corresponds to a B$_{hcp}$N$_{top}$ registry, while the regions on the wire with a B$_{fcc}$N$_{hcp}$ registry have the weakest interaction with the substrate. The $p_{z}$ orbitals of the nitrogen atoms partially hybridize with the metal substrate. As the magnitude of the hybridization depends on the interaction strength and thus on the atomic registry, it gives rise to a modulation of the work function within the moir\'e unit cell of $\sim$0.5 eV. Overall, the h-BN layer reduces the work function of the Ir(111) substrate by more than 1 eV. Our results are in line with previous findings that the strength of the chemical interaction at the interface of h-BN and Ir(111) should be less than for Rh(111) or Ru(0001) but more than for Pt(111).\cite{Preo2007} Therefore, the moir\'e pattern formed by h-BN/Ir(111) combines the advantages found in strongly interacting h-BN/metal system of large work function modulation and single domain growth with a low structural corrugation, the latter usually characteristic for weakly interacting h-BN/metal systems. This makes it a superior candidate for the bottom-up fabrication of electronically modulated thin films and hetero-sandwich structures.

\begin{acknowledgments}
This research made use of the Aalto Nanomicroscopy Center (Aalto NMC) facilities and was supported by the European Research Council (ERC-2011-StG No. 278698 "PRECISE-NANO"), the Academy of Finland (Centre of Excellence in Low Temperature Quantum Phenomena and Devices No. 250280), and the Finnish Academy of Science and Letters.
\end{acknowledgments}


\begin{thebibliography}{10}%
\makeatletter
\providecommand \@ifxundefined [1]{%
 \ifx #1\undefined \expandafter \@firstoftwo
 \else \expandafter \@secondoftwo
\fi
}%
\providecommand \@ifnum [1]{%
 \ifnum #1\expandafter \@firstoftwo
 \else \expandafter \@secondoftwo
\fi
}%
\providecommand \enquote [1]{``#1''}%
\providecommand \bibnamefont  [1]{#1}%
\providecommand \bibfnamefont [1]{#1}%
\providecommand \citenamefont [1]{#1}%
\providecommand\href[0]{\@sanitize\@href}%
\providecommand\@href[1]{\endgroup\@@startlink{#1}\endgroup\@@href}%
\providecommand\@@href[1]{#1\@@endlink}%
\providecommand \@sanitize [0]{\begingroup\catcode`\&12\catcode`\#12\relax}%
\@ifxundefined \pdfoutput {\@firstoftwo}{%
 \@ifnum{\z@=\pdfoutput}{\@firstoftwo}{\@secondoftwo}%
}{%
 \providecommand\@@startlink[1]{\leavevmode}%
 \providecommand\@@endlink[0]{}%
}{%
 \providecommand\@@startlink[1]{%
  \leavevmode
  \pdfstartlink
   attr{/Border[0 0 1 ]/H/I/C[0 1 1]}%
   user{/Subtype/Link/A<</Type/Action/S/URI/URI(#1)>>}%
  \relax
 }%
 \providecommand\@@endlink[0]{\pdfendlink}%
}%
\providecommand \url  [0]{\begingroup\@sanitize \@url }%
\providecommand \@url [1]{\endgroup\@href {#1}{\urlprefix}}%
\providecommand \urlprefix [0]{URL }%
\providecommand \Eprint[0]{\href }%
\@ifxundefined \urlstyle {%
  \providecommand \doi [1]{doi:\discretionary{}{}{}#1}%
}{%
  \providecommand \doi [0]{doi:\discretionary{}{}{}\begingroup
  \urlstyle{rm}\Url }%
}%
\providecommand \doibase [0]{http://dx.doi.org/}%
\providecommand \Doi[1]{\href{\doibase#1}}%
\providecommand \bibAnnote [3]{%
  \BibitemShut{#1}%
  \begin{quotation}\noindent
    \textsc{Key:}\ #2\\\textsc{Annotation:}\ #3%
  \end{quotation}%
}%
\providecommand \bibAnnoteFile [2]{%
  \IfFileExists{#2}{\bibAnnote {#1} {#2} {\input{#2}}}{}%
}%
\providecommand \typeout [0]{\immediate \write \m@ne }%
\providecommand \selectlanguage [0]{\@gobble}%
\providecommand \bibinfo [0]{\@secondoftwo}%
\providecommand \bibfield [0]{\@secondoftwo}%
\providecommand \translation [1]{[#1]}%
\providecommand \BibitemOpen[0]{}%
\providecommand \bibitemStop [0]{}%
\providecommand \bibitemNoStop [0]{.\EOS\space}%
\providecommand \EOS [0]{\spacefactor3000\relax}%
\providecommand \BibitemShut [1]{\csname bibitem#1\endcsname}%
\bibitem{Dean2010}%
  \BibitemOpen
  \bibfield{author}{%
  \bibinfo {author} {\bibfnamefont{C.~R.}\ \bibnamefont{Dean}}, \bibinfo
  {author} {\bibfnamefont{A.~F.}\ \bibnamefont{Young}}, \bibinfo {author}
  {\bibfnamefont{I.}~\bibnamefont{Meric}}, \bibinfo {author}
  {\bibfnamefont{C.}~\bibnamefont{Lee}}, \bibinfo {author}
  {\bibfnamefont{L.}~\bibnamefont{Wang}}, \bibinfo {author}
  {\bibfnamefont{S.}~\bibnamefont{Sorgenfrei}}, \bibinfo {author}
  {\bibfnamefont{K.}~\bibnamefont{Watanabe}}, \bibinfo {author}
  {\bibfnamefont{T.}~\bibnamefont{Taniguchi}}, \bibinfo {author}
  {\bibfnamefont{P.}~\bibnamefont{Kim}}, \bibinfo {author}
  {\bibfnamefont{K.~L.}\ \bibnamefont{Shepard}},\ and\ \bibinfo {author}
  {\bibfnamefont{J.}~\bibnamefont{Hone}},\ }%
  \bibfield{journal}{%
  \bibinfo {journal} {Nat. Nanotechnol.}\ }%
  \textbf{\bibinfo {volume} {5}},\ \bibinfo {pages} {722} (\bibinfo {year}
  {2010})%
  \bibAnnoteFile{NoStop}{Dean2010}%
\bibitem{Hunt2013}%
  \BibitemOpen
  \bibfield{author}{%
  \bibinfo {author} {\bibfnamefont{B.}~\bibnamefont{Hunt}}, \bibinfo {author}
  {\bibfnamefont{J.~D.}\ \bibnamefont{Sanchez-Yamagishi}}, \bibinfo {author}
  {\bibfnamefont{A.~F.}\ \bibnamefont{Young}}, \bibinfo {author}
  {\bibfnamefont{M.}~\bibnamefont{Yankowitz}}, \bibinfo {author}
  {\bibfnamefont{B.~J.}\ \bibnamefont{LeRoy}}, \bibinfo {author}
  {\bibfnamefont{K.}~\bibnamefont{Watanabe}}, \bibinfo {author}
  {\bibfnamefont{T.}~\bibnamefont{Taniguchi}}, \bibinfo {author}
  {\bibfnamefont{P.}~\bibnamefont{Moon}}, \bibinfo {author}
  {\bibfnamefont{M.}~\bibnamefont{Koshino}}, \bibinfo {author}
  {\bibfnamefont{P.}~\bibnamefont{Jarillo-Herrero}},\ and\ \bibinfo {author}
  {\bibfnamefont{R.~C.}\ \bibnamefont{Ashoori}},\ }%
  \bibfield{journal}{%
  \bibinfo {journal} {Science}\ }%
  \textbf{\bibinfo {volume} {340}},\ \bibinfo {pages} {1427} (\bibinfo {year}
  {2013})%
  \bibAnnoteFile{NoStop}{Hunt2013}%
\bibitem{Liu2014}%
  \BibitemOpen
  \bibfield{author}{%
  \bibinfo {author} {\bibfnamefont{L.}~\bibnamefont{Liu}}, \bibinfo {author}
  {\bibfnamefont{J.}~\bibnamefont{Park}}, \bibinfo {author}
  {\bibfnamefont{D.~A.}\ \bibnamefont{Siegel}}, \bibinfo {author}
  {\bibfnamefont{K.~F.}\ \bibnamefont{McCarty}}, \bibinfo {author}
  {\bibfnamefont{K.~W.}\ \bibnamefont{Clark}}, \bibinfo {author}
  {\bibfnamefont{W.}~\bibnamefont{Deng}}, \bibinfo {author}
  {\bibfnamefont{L.}~\bibnamefont{Basile}}, \bibinfo {author}
  {\bibfnamefont{J.~C.}\ \bibnamefont{Idrobo}}, \bibinfo {author}
  {\bibfnamefont{A.-P.}\ \bibnamefont{Li}},\ and\ \bibinfo {author}
  {\bibfnamefont{G.}~\bibnamefont{Gu}},\ }%
  \bibfield{journal}{%
  \bibinfo {journal} {Science}\ }%
  \textbf{\bibinfo {volume} {343}},\ \bibinfo {pages} {163} (\bibinfo {year}
  {2014})%
  \bibAnnoteFile{NoStop}{Liu2014}%
\bibitem{Corso2004}%
  \BibitemOpen
  \bibfield{author}{%
  \bibinfo {author} {\bibfnamefont{M.}~\bibnamefont{Corso}}, \bibinfo {author}
  {\bibfnamefont{W.}~\bibnamefont{Auw\"arter}}, \bibinfo {author}
  {\bibfnamefont{M.}~\bibnamefont{Muntwiler}}, \bibinfo {author}
  {\bibfnamefont{A.}~\bibnamefont{Tamai}}, \bibinfo {author}
  {\bibfnamefont{T.}~\bibnamefont{Greber}},\ and\ \bibinfo {author}
  {\bibfnamefont{J.}~\bibnamefont{Osterwalder}},\ }%
  \bibfield{journal}{%
  \bibinfo {journal} {Science}\ }%
  \textbf{\bibinfo {volume} {303}},\ \bibinfo {pages} {217} (\bibinfo {year}
  {2004})%
  \bibAnnoteFile{NoStop}{Corso2004}%
\bibitem{Laskowski2007}%
  \BibitemOpen
  \bibfield{author}{%
  \bibinfo {author} {\bibfnamefont{R.}~\bibnamefont{Laskowski}}, \bibinfo
  {author} {\bibfnamefont{P.}~\bibnamefont{Blaha}}, \bibinfo {author}
  {\bibfnamefont{T.}~\bibnamefont{Gallauner}},\ and\ \bibinfo {author}
  {\bibfnamefont{K.}~\bibnamefont{Schwarz}},\ }%
  \bibfield{journal}{%
  \bibinfo {journal} {Phys. Rev. Lett.}\ }%
  \textbf{\bibinfo {volume} {98}},\ \bibinfo {pages} {106802} (\bibinfo {year}
  {2007})%
  \bibAnnoteFile{NoStop}{Laskowski2007}%
\bibitem{Corso2005}%
  \BibitemOpen
  \bibfield{author}{%
  \bibinfo {author} {\bibfnamefont{M.}~\bibnamefont{Corso}}, \bibinfo {author}
  {\bibfnamefont{T.}~\bibnamefont{Greber}},\ and\ \bibinfo {author}
  {\bibfnamefont{J.}~\bibnamefont{Osterwalder}},\ }%
  \bibfield{journal}{%
  \bibinfo {journal} {Surf. Sci.}\ }%
  \textbf{\bibinfo {volume} {577}},\ \bibinfo {pages} {L78} (\bibinfo {year}
  {2005})%
  \bibAnnoteFile{NoStop}{Corso2005}%
\bibitem{Goriachko2007}%
  \BibitemOpen
  \bibfield{author}{%
  \bibinfo {author} {\bibfnamefont{A.}~\bibnamefont{Goriachko}}, \bibinfo
  {author} {\bibfnamefont{Y.}~\bibnamefont{He}}, \bibinfo {author}
  {\bibfnamefont{M.}~\bibnamefont{Knapp}}, \bibinfo {author}
  {\bibfnamefont{H.}~\bibnamefont{Over}}, \bibinfo {author}
  {\bibfnamefont{M.}~\bibnamefont{Corso}}, \bibinfo {author}
  {\bibfnamefont{T.}~\bibnamefont{Brugger}}, \bibinfo {author}
  {\bibfnamefont{S.}~\bibnamefont{Berner}}, \bibinfo {author}
  {\bibfnamefont{J.}~\bibnamefont{Osterwalder}},\ and\ \bibinfo {author}
  {\bibfnamefont{T.}~\bibnamefont{Greber}},\ }%
  \bibfield{journal}{%
  \bibinfo {journal} {Langmuir}\ }%
  \textbf{\bibinfo {volume} {23}},\ \bibinfo {pages} {2928} (\bibinfo {year}
  {2007})%
  \bibAnnoteFile{NoStop}{Goriachko2007}%
\bibitem{Berner2007}%
  \BibitemOpen
  \bibfield{author}{%
  \bibinfo {author} {\bibfnamefont{S.}~\bibnamefont{Berner}}, \bibinfo {author}
  {\bibfnamefont{M.}~\bibnamefont{Corso}}, \bibinfo {author}
  {\bibfnamefont{R.}~\bibnamefont{Widmer}}, \bibinfo {author}
  {\bibfnamefont{O.}~\bibnamefont{Groening}}, \bibinfo {author}
  {\bibfnamefont{R.}~\bibnamefont{Laskowski}}, \bibinfo {author}
  {\bibfnamefont{P.}~\bibnamefont{Blaha}}, \bibinfo {author}
  {\bibfnamefont{K.}~\bibnamefont{Schwarz}}, \bibinfo {author}
  {\bibfnamefont{A.}~\bibnamefont{Goriachko}}, \bibinfo {author}
  {\bibfnamefont{H.}~\bibnamefont{Over}}, \bibinfo {author}
  {\bibfnamefont{S.}~\bibnamefont{Gsell}}, \bibinfo {author}
  {\bibfnamefont{M.}~\bibnamefont{Schreck}}, \bibinfo {author}
  {\bibfnamefont{H.}~\bibnamefont{Sachdev}}, \bibinfo {author}
  {\bibfnamefont{T.}~\bibnamefont{Greber}},\ and\ \bibinfo {author}
  {\bibfnamefont{J.}~\bibnamefont{Osterwalder}},\ }%
  \bibfield{journal}{%
  \bibinfo {journal} {Angew. Chem. Int. Ed.}\ }%
  \textbf{\bibinfo {volume} {46}},\ \bibinfo {pages} {5115} (\bibinfo {year}
  {2007})%
  \bibAnnoteFile{NoStop}{Berner2007}%
\bibitem{Dil2008}%
  \BibitemOpen
  \bibfield{author}{%
  \bibinfo {author} {\bibfnamefont{H.}~\bibnamefont{Dil}}, \bibinfo {author}
  {\bibfnamefont{J.}~\bibnamefont{Lobo-Checa}}, \bibinfo {author}
  {\bibfnamefont{R.}~\bibnamefont{Laskowski}}, \bibinfo {author}
  {\bibfnamefont{P.}~\bibnamefont{Blaha}}, \bibinfo {author}
  {\bibfnamefont{S.}~\bibnamefont{Berner}}, \bibinfo {author}
  {\bibfnamefont{J.}~\bibnamefont{Osterwalder}},\ and\ \bibinfo {author}
  {\bibfnamefont{T.}~\bibnamefont{Greber}},\ }%
  \bibfield{journal}{%
  \bibinfo {journal} {Science}\ }%
  \textbf{\bibinfo {volume} {319}},\ \bibinfo {pages} {1824} (\bibinfo {year}
  {2008})%
  \bibAnnoteFile{NoStop}{Dil2008}%
\bibitem{Kahle2012}%
  \BibitemOpen
  \bibfield{author}{%
  \bibinfo {author} {\bibfnamefont{S.}~\bibnamefont{Kahle}}, \bibinfo {author}
  {\bibfnamefont{Z.}~\bibnamefont{Deng}}, \bibinfo {author}
  {\bibfnamefont{N.}~\bibnamefont{Malinowski}}, \bibinfo {author}
  {\bibfnamefont{C.}~\bibnamefont{Tonnoir}}, \bibinfo {author}
  {\bibfnamefont{A.}~\bibnamefont{Forment-Aliaga}}, \bibinfo {author}
  {\bibfnamefont{N.}~\bibnamefont{Thontasen}}, \bibinfo {author}
  {\bibfnamefont{G.}~\bibnamefont{Rinke}}, \bibinfo {author}
  {\bibfnamefont{D.}~\bibnamefont{Le}}, \bibinfo {author}
  {\bibfnamefont{V.}~\bibnamefont{Turkowski}}, \bibinfo {author}
  {\bibfnamefont{T.~S.}\ \bibnamefont{Rahman}}, \bibinfo {author}
  {\bibfnamefont{S.}~\bibnamefont{Rauschenbach}}, \bibinfo {author}
  {\bibfnamefont{M.}~\bibnamefont{Ternes}},\ and\ \bibinfo {author}
  {\bibfnamefont{K.}~\bibnamefont{Kern}},\ }%
  \bibfield{journal}{%
  \bibinfo {journal} {Nano Lett.}\ }%
  \textbf{\bibinfo {volume} {12}},\ \bibinfo {pages} {518} (\bibinfo {year}
  {2012})%
  \bibAnnoteFile{NoStop}{Kahle2012}%
\bibitem{Joshi2012}%
  \BibitemOpen
  \bibfield{author}{%
  \bibinfo {author} {\bibfnamefont{S.}~\bibnamefont{Joshi}}, \bibinfo {author}
  {\bibfnamefont{D.}~\bibnamefont{Ecija}}, \bibinfo {author}
  {\bibfnamefont{R.}~\bibnamefont{Koitz}}, \bibinfo {author}
  {\bibfnamefont{M.}~\bibnamefont{Iannuzzi}}, \bibinfo {author}
  {\bibfnamefont{A.~P.}\ \bibnamefont{Seitsonen}}, \bibinfo {author}
  {\bibfnamefont{J.}~\bibnamefont{Hutter}}, \bibinfo {author}
  {\bibfnamefont{H.}~\bibnamefont{Sachdev}}, \bibinfo {author}
  {\bibfnamefont{S.}~\bibnamefont{Vijayaraghavan}}, \bibinfo {author}
  {\bibfnamefont{F.}~\bibnamefont{Bischoff}}, \bibinfo {author}
  {\bibfnamefont{K.}~\bibnamefont{Seufert}}, \bibinfo {author}
  {\bibfnamefont{J.~V.}\ \bibnamefont{Barth}},\ and\ \bibinfo {author}
  {\bibfnamefont{W.}~\bibnamefont{Auwaerter}},\ }%
  \bibfield{journal}{%
  \bibinfo {journal} {Nano Lett.}\ }%
  \textbf{\bibinfo {volume} {12}},\ \bibinfo {pages} {5821} (\bibinfo {year}
  {2012})%
  \bibAnnoteFile{NoStop}{Joshi2012}%
\bibitem{Schulz2013}%
  \BibitemOpen
  \bibfield{author}{%
  \bibinfo {author} {\bibfnamefont{F.}~\bibnamefont{Schulz}}, \bibinfo {author}
  {\bibfnamefont{R.}~\bibnamefont{Drost}}, \bibinfo {author}
  {\bibfnamefont{S.~K.}\ \bibnamefont{H\"am\"al\"ainen}},\ and\ \bibinfo
  {author} {\bibfnamefont{P.}~\bibnamefont{Liljeroth}},\ }%
  \bibfield{journal}{%
  \bibinfo {journal} {ACS Nano}\ }%
  \textbf{\bibinfo {volume} {7}},\ \bibinfo {pages} {11121} (\bibinfo {year}
  {2013})%
  \bibAnnoteFile{NoStop}{Schulz2013}%
\bibitem{Joshi2013}%
  \BibitemOpen
  \bibfield{author}{%
  \bibinfo {author} {\bibfnamefont{S.}~\bibnamefont{Joshi}}, \bibinfo {author}
  {\bibfnamefont{F.}~\bibnamefont{Bischoff}}, \bibinfo {author}
  {\bibfnamefont{R.}~\bibnamefont{Koitz}}, \bibinfo {author}
  {\bibfnamefont{D.}~\bibnamefont{Ecija}}, \bibinfo {author}
  {\bibfnamefont{K.}~\bibnamefont{Seufert}}, \bibinfo {author}
  {\bibfnamefont{A.~P.}\ \bibnamefont{Seitsonen}}, \bibinfo {author}
  {\bibfnamefont{J.}~\bibnamefont{Hutter}}, \bibinfo {author}
  {\bibfnamefont{K.}~\bibnamefont{Diller}}, \bibinfo {author}
  {\bibfnamefont{J.~I.}\ \bibnamefont{Urgel}}, \bibinfo {author}
  {\bibfnamefont{H.}~\bibnamefont{Sachdev}}, \bibinfo {author}
  {\bibfnamefont{J.~V.}\ \bibnamefont{Barth}},\ and\ \bibinfo {author}
  {\bibfnamefont{W.}~\bibnamefont{Auw\"arter}},\ }%
  \bibfield{journal}{%
  \bibinfo {journal} {ACS Nano}\ }%
  \textbf{\bibinfo {volume} {8}},\ \bibinfo {pages} {430} (\bibinfo {year}
  {2014})%
  \bibAnnoteFile{NoStop}{Joshi2013}%
\bibitem{Preo2005}%
  \BibitemOpen
  \bibfield{author}{%
  \bibinfo {author} {\bibfnamefont{A.}~\bibnamefont{Preobrajenski}}, \bibinfo
  {author} {\bibfnamefont{A.}~\bibnamefont{Vinogradov}},\ and\ \bibinfo
  {author} {\bibfnamefont{N.}~\bibnamefont{M\r{a}rtensson}},\ }%
  \bibfield{journal}{%
  \bibinfo {journal} {Surf. Sci.}\ }%
  \textbf{\bibinfo {volume} {582}},\ \bibinfo {pages} {21} (\bibinfo {year}
  {2005})%
  \bibAnnoteFile{NoStop}{Preo2005}%
\bibitem{Roth2013}%
  \BibitemOpen
  \bibfield{author}{%
  \bibinfo {author} {\bibfnamefont{S.}~\bibnamefont{Roth}}, \bibinfo {author}
  {\bibfnamefont{F.}~\bibnamefont{Matsui}}, \bibinfo {author}
  {\bibfnamefont{T.}~\bibnamefont{Greber}},\ and\ \bibinfo {author}
  {\bibfnamefont{J.}~\bibnamefont{Osterwalder}},\ }%
  \bibfield{journal}{%
  \bibinfo {journal} {Nano Lett.}\ }%
  \textbf{\bibinfo {volume} {13}},\ \bibinfo {pages} {2668} (\bibinfo {year}
  {2013})%
  \bibAnnoteFile{NoStop}{Roth2013}%
\bibitem{Preo2007}%
  \BibitemOpen
  \bibfield{author}{%
  \bibinfo {author} {\bibfnamefont{A.~B.}\ \bibnamefont{Preobrajenski}},
  \bibinfo {author} {\bibfnamefont{M.~A.}\ \bibnamefont{Nesterov}}, \bibinfo
  {author} {\bibfnamefont{M.~L.}\ \bibnamefont{Ng}}, \bibinfo {author}
  {\bibfnamefont{A.~S.}\ \bibnamefont{Vinogradov}},\ and\ \bibinfo {author}
  {\bibfnamefont{N.}~\bibnamefont{Martensson}},\ }%
  \bibfield{journal}{%
  \bibinfo {journal} {Chem. Phys. Lett.}\ }%
  \textbf{\bibinfo {volume} {446}},\ \bibinfo {pages} {119} (\bibinfo {year}
  {2007})%
  \bibAnnoteFile{NoStop}{Preo2007}%
\bibitem{Mueller2005}%
  \BibitemOpen
  \bibfield{author}{%
  \bibinfo {author} {\bibfnamefont{F.}~\bibnamefont{M\"uller}}, \bibinfo
  {author} {\bibfnamefont{K.}~\bibnamefont{St\"owe}},\ and\ \bibinfo {author}
  {\bibfnamefont{H.}~\bibnamefont{Sachdev}},\ }%
  \bibfield{journal}{%
  \bibinfo {journal} {Chem. Mater.}\ }%
  \textbf{\bibinfo {volume} {17}},\ \bibinfo {pages} {3464} (\bibinfo {year}
  {2005})%
  \bibAnnoteFile{NoStop}{Mueller2005}%
\bibitem{Morscher2006}%
  \BibitemOpen
  \bibfield{author}{%
  \bibinfo {author} {\bibfnamefont{M.}~\bibnamefont{Morscher}}, \bibinfo
  {author} {\bibfnamefont{M.}~\bibnamefont{Corso}}, \bibinfo {author}
  {\bibfnamefont{T.}~\bibnamefont{Greber}},\ and\ \bibinfo {author}
  {\bibfnamefont{J.}~\bibnamefont{Osterwalder}},\ }%
  \bibfield{journal}{%
  \bibinfo {journal} {Surf. Sci.}\ }%
  \textbf{\bibinfo {volume} {600}},\ \bibinfo {pages} {3280} (\bibinfo {year}
  {2006})%
  \bibAnnoteFile{NoStop}{Morscher2006}%
\bibitem{Mueller2010}%
  \BibitemOpen
  \bibfield{author}{%
  \bibinfo {author} {\bibfnamefont{F.}~\bibnamefont{M\"uller}}, \bibinfo
  {author} {\bibfnamefont{S.}~\bibnamefont{H\"ufner}}, \bibinfo {author}
  {\bibfnamefont{H.}~\bibnamefont{Sachdev}}, \bibinfo {author}
  {\bibfnamefont{R.}~\bibnamefont{Laskowski}}, \bibinfo {author}
  {\bibfnamefont{P.}~\bibnamefont{Blaha}},\ and\ \bibinfo {author}
  {\bibfnamefont{K.}~\bibnamefont{Schwarz}},\ }%
  \bibfield{journal}{%
  \bibinfo {journal} {Phys. Rev. B}\ }%
  \textbf{\bibinfo {volume} {82}},\ \bibinfo {pages} {113406} (\bibinfo {year}
  {2010})%
  \bibAnnoteFile{NoStop}{Mueller2010}%
\bibitem{Diaz2013}%
  \BibitemOpen
  \bibfield{author}{%
  \bibinfo {author} {\bibfnamefont{J.~G.}\ \bibnamefont{Diaz}}, \bibinfo
  {author} {\bibfnamefont{Y.}~\bibnamefont{Ding}}, \bibinfo {author}
  {\bibfnamefont{R.}~\bibnamefont{Koitz}}, \bibinfo {author}
  {\bibfnamefont{A.~P.}\ \bibnamefont{Seitsonen}}, \bibinfo {author}
  {\bibfnamefont{M.}~\bibnamefont{Iannuzzi}},\ and\ \bibinfo {author}
  {\bibfnamefont{J.}~\bibnamefont{Hutter}},\ }%
  \bibfield{journal}{%
  \bibinfo {journal} {Theor. Chem. Acc.}\ }%
  \textbf{\bibinfo {volume} {132}},\ \bibinfo {pages} {1350} (\bibinfo {year}
  {2013})%
  \bibAnnoteFile{NoStop}{Diaz2013}%
\bibitem{Orlando2012}%
  \BibitemOpen
  \bibfield{author}{%
  \bibinfo {author} {\bibfnamefont{F.}~\bibnamefont{Orlando}}, \bibinfo
  {author} {\bibfnamefont{R.}~\bibnamefont{Larciprete}}, \bibinfo {author}
  {\bibfnamefont{P.}~\bibnamefont{Lacovig}}, \bibinfo {author}
  {\bibfnamefont{I.}~\bibnamefont{Boscarato}}, \bibinfo {author}
  {\bibfnamefont{A.}~\bibnamefont{Baraldi}},\ and\ \bibinfo {author}
  {\bibfnamefont{S.}~\bibnamefont{Lizzit}},\ }%
  \bibfield{journal}{%
  \bibinfo {journal} {J. Phys. Chem. C}\ }%
  \textbf{\bibinfo {volume} {116}},\ \bibinfo {pages} {157} (\bibinfo {year}
  {2012})%
  \bibAnnoteFile{NoStop}{Orlando2012}%
\bibitem{Laskowski2008}%
  \BibitemOpen
  \bibfield{author}{%
  \bibinfo {author} {\bibfnamefont{R.}~\bibnamefont{Laskowski}}, \bibinfo
  {author} {\bibfnamefont{P.}~\bibnamefont{Blaha}},\ and\ \bibinfo {author}
  {\bibfnamefont{K.}~\bibnamefont{Schwarz}},\ }%
  \bibfield{journal}{%
  \bibinfo {journal} {Phys. Rev. B}\ }%
  \textbf{\bibinfo {volume} {78}},\ \bibinfo {pages} {045409} (\bibinfo {year}
  {2008})%
  \bibAnnoteFile{NoStop}{Laskowski2008}%
\bibitem{Lu2013}%
  \BibitemOpen
  \bibfield{author}{%
  \bibinfo {author} {\bibfnamefont{J.}~\bibnamefont{Lu}}, \bibinfo {author}
  {\bibfnamefont{P.~S.~E.}\ \bibnamefont{Yeo}}, \bibinfo {author}
  {\bibfnamefont{Y.}~\bibnamefont{Zheng}}, \bibinfo {author}
  {\bibfnamefont{H.}~\bibnamefont{Xu}}, \bibinfo {author}
  {\bibfnamefont{C.~K.}\ \bibnamefont{Gan}}, \bibinfo {author}
  {\bibfnamefont{M.~B.}\ \bibnamefont{Sullivan}}, \bibinfo {author}
  {\bibfnamefont{A.}~\bibnamefont{Castro~Neto}},\ and\ \bibinfo {author}
  {\bibfnamefont{K.~P.}\ \bibnamefont{Loh}},\ }%
  \bibfield{journal}{%
  \bibinfo {journal} {J. Am. Chem. Soc.}\ }%
  \textbf{\bibinfo {volume} {135}},\ \bibinfo {pages} {2368} (\bibinfo {year}
  {2013})%
  \bibAnnoteFile{NoStop}{Lu2013}%
\bibitem{Horcas2007}%
  \BibitemOpen
  \bibfield{author}{%
  \bibinfo {author} {\bibfnamefont{I.}~\bibnamefont{Horcas}}, \bibinfo {author}
  {\bibfnamefont{R.}~\bibnamefont{Fernandez}}, \bibinfo {author}
  {\bibfnamefont{J.~M.}\ \bibnamefont{Gomez-Rodriguez}}, \bibinfo {author}
  {\bibfnamefont{J.}~\bibnamefont{Colchero}}, \bibinfo {author}
  {\bibfnamefont{J.}~\bibnamefont{Gomez-Herrero}},\ and\ \bibinfo {author}
  {\bibfnamefont{A.~M.}\ \bibnamefont{Baro}},\ }%
  \bibfield{journal}{%
  \bibinfo {journal} {Rev. Sci. Instrum.}\ }%
  \textbf{\bibinfo {volume} {78}},\ \bibinfo {pages} {013705} (\bibinfo {year}
  {2007})%
  \bibAnnoteFile{NoStop}{Horcas2007}%
\bibitem{SpectraFox}%
  \BibitemOpen
  \bibfield{author}{%
  \bibinfo {author} {\bibfnamefont{M.}~\bibnamefont{Ruby}},\ }%
  \emph{\bibinfo {title} {SpectraFox}},\ \bibinfo {howpublished}
  {\url{http://www.spectrafox.com}}%
  \bibAnnoteFile{NoStop}{SpectraFox}%
\bibitem{VandeVondele2005}%
  \BibitemOpen
  \bibfield{author}{%
  \bibinfo {author} {\bibfnamefont{J.}~\bibnamefont{VandeVondele}}, \bibinfo
  {author} {\bibfnamefont{M.}~\bibnamefont{Krack}}, \bibinfo {author}
  {\bibfnamefont{F.}~\bibnamefont{Mohamed}}, \bibinfo {author}
  {\bibfnamefont{M.}~\bibnamefont{Parrinello}}, \bibinfo {author}
  {\bibfnamefont{T.}~\bibnamefont{Chassaing}},\ and\ \bibinfo {author}
  {\bibfnamefont{J.}~\bibnamefont{Hutter}},\ }%
  \bibfield{journal}{%
  \bibinfo {journal} {Comput. Phys. Commun.}\ }%
  \textbf{\bibinfo {volume} {167}},\ \bibinfo {pages} {103} (\bibinfo {year}
  {2005})%
  \bibAnnoteFile{NoStop}{VandeVondele2005}%
\bibitem{VandeVondele2007}%
  \BibitemOpen
  \bibfield{author}{%
  \bibinfo {author} {\bibfnamefont{J.}~\bibnamefont{VandeVondele}}\ and\
  \bibinfo {author} {\bibfnamefont{J.}~\bibnamefont{Hutter}},\ }%
  \bibfield{journal}{%
  \bibinfo {journal} {J. Chem. Phys.}\ }%
  \textbf{\bibinfo {volume} {127}},\ \bibinfo {pages} {114105} (\bibinfo {year}
  {2007})%
  \bibAnnoteFile{NoStop}{VandeVondele2007}%
\bibitem{Zhang1998}%
  \BibitemOpen
  \bibfield{author}{%
  \bibinfo {author} {\bibfnamefont{Y.~K.}\ \bibnamefont{Zhang}}\ and\ \bibinfo
  {author} {\bibfnamefont{W.~T.}\ \bibnamefont{Yang}},\ }%
  \bibfield{journal}{%
  \bibinfo {journal} {Phys. Rev. Lett.}\ }%
  \textbf{\bibinfo {volume} {80}},\ \bibinfo {pages} {890} (\bibinfo {year}
  {1998})%
  \bibAnnoteFile{NoStop}{Zhang1998}%
\bibitem{Grimme2010}%
  \BibitemOpen
  \bibfield{author}{%
  \bibinfo {author} {\bibfnamefont{S.}~\bibnamefont{Grimme}}, \bibinfo {author}
  {\bibfnamefont{J.}~\bibnamefont{Antony}}, \bibinfo {author}
  {\bibfnamefont{S.}~\bibnamefont{Ehrlich}},\ and\ \bibinfo {author}
  {\bibfnamefont{H.}~\bibnamefont{Krieg}},\ }%
  \bibfield{journal}{%
  \bibinfo {journal} {J. Chem. Phys.}\ }%
  \textbf{\bibinfo {volume} {132}},\ \bibinfo {pages} {154104} (\bibinfo {year}
  {2010})%
  \bibAnnoteFile{NoStop}{Grimme2010}%
\bibitem{Tersoff1985}%
  \BibitemOpen
  \bibfield{author}{%
  \bibinfo {author} {\bibfnamefont{J.}~\bibnamefont{Tersoff}}\ and\ \bibinfo
  {author} {\bibfnamefont{D.~R.}\ \bibnamefont{Hamann}},\ }%
  \bibfield{journal}{%
  \bibinfo {journal} {Phys. Rev. B}\ }%
  \textbf{\bibinfo {volume} {31}},\ \bibinfo {pages} {805} (\bibinfo {year}
  {1985})%
  \bibAnnoteFile{NoStop}{Tersoff1985}%
\bibitem{Coraux2008}%
  \BibitemOpen
  \bibfield{author}{%
  \bibinfo {author} {\bibfnamefont{J.}~\bibnamefont{Coraux}}, \bibinfo {author}
  {\bibfnamefont{A.~T.}\ \bibnamefont{N‘Diaye}}, \bibinfo {author}
  {\bibfnamefont{C.}~\bibnamefont{Busse}},\ and\ \bibinfo {author}
  {\bibfnamefont{T.}~\bibnamefont{Michely}},\ }%
  \bibfield{journal}{%
  \bibinfo {journal} {Nano Letters}\ }%
  \textbf{\bibinfo {volume} {8}},\ \bibinfo {pages} {565} (\bibinfo {year}
  {2008})%
  \bibAnnoteFile{NoStop}{Coraux2008}%
\bibitem{Sutter2008}%
  \BibitemOpen
  \bibfield{author}{%
  \bibinfo {author} {\bibfnamefont{P.~W.}\ \bibnamefont{Sutter}}, \bibinfo
  {author} {\bibfnamefont{J.-I.}\ \bibnamefont{Flege}},\ and\ \bibinfo {author}
  {\bibfnamefont{E.~A.}\ \bibnamefont{Sutter}},\ }%
  \bibfield{journal}{%
  \bibinfo {journal} {Nature Materials}\ }%
  \textbf{\bibinfo {volume} {7}},\ \bibinfo {pages} {406} (\bibinfo {year}
  {2008})%
  \bibAnnoteFile{NoStop}{Sutter2008}%
\bibitem{Singh1968}%
  \BibitemOpen
  \bibfield{author}{%
  \bibinfo {author} {\bibfnamefont{H.~P.}\ \bibnamefont{Singh}},\ }%
  \bibfield{journal}{%
  \bibinfo {journal} {Acta Crystallogr. Sect. A}\ }%
  \textbf{\bibinfo {volume} {24}},\ \bibinfo {pages} {469} (\bibinfo {year}
  {1968})%
  \bibAnnoteFile{NoStop}{Singh1968}%
\bibitem{Pease1952}%
  \BibitemOpen
  \bibfield{author}{%
  \bibinfo {author} {\bibfnamefont{R.~S.}\ \bibnamefont{Pease}},\ }%
  \bibfield{journal}{%
  \bibinfo {journal} {Acta Crystallogr.}\ }%
  \textbf{\bibinfo {volume} {5}},\ \bibinfo {pages} {356} (\bibinfo {year}
  {1952})%
  \bibAnnoteFile{NoStop}{Pease1952}%
\bibitem{NDiaye2008}%
  \BibitemOpen
  \bibfield{author}{%
  \bibinfo {author} {\bibfnamefont{A.~T.}\ \bibnamefont{N'Diaye}}, \bibinfo
  {author} {\bibfnamefont{J.}~\bibnamefont{Coraux}}, \bibinfo {author}
  {\bibfnamefont{T.~N.}\ \bibnamefont{Plasa}}, \bibinfo {author}
  {\bibfnamefont{C.}~\bibnamefont{Busse}},\ and\ \bibinfo {author}
  {\bibfnamefont{T.}~\bibnamefont{Michely}},\ }%
  \bibfield{journal}{%
  \bibinfo {journal} {New J. Phys.}\ }%
  \textbf{\bibinfo {volume} {10}},\ \bibinfo {pages} {043033} (\bibinfo {year}
  {2008})%
  \bibAnnoteFile{NoStop}{NDiaye2008}%
\bibitem{Blanc2012}%
  \BibitemOpen
  \bibfield{author}{%
  \bibinfo {author} {\bibfnamefont{N.}~\bibnamefont{Blanc}}, \bibinfo {author}
  {\bibfnamefont{J.}~\bibnamefont{Coraux}}, \bibinfo {author}
  {\bibfnamefont{C.}~\bibnamefont{Vo-Van}}, \bibinfo {author}
  {\bibfnamefont{A.~T.}\ \bibnamefont{N'Diaye}}, \bibinfo {author}
  {\bibfnamefont{O.}~\bibnamefont{Geaymond}},\ and\ \bibinfo {author}
  {\bibfnamefont{G.}~\bibnamefont{Renaud}},\ }%
  \bibfield{journal}{%
  \bibinfo {journal} {Phys. Rev. B}\ }%
  \textbf{\bibinfo {volume} {86}},\ \bibinfo {pages} {235439} (\bibinfo {month}
  {Dec}\ \bibinfo {year} {2012})%
  \bibAnnoteFile{NoStop}{Blanc2012}%
\bibitem{Hattab2012}%
  \BibitemOpen
  \bibfield{author}{%
  \bibinfo {author} {\bibfnamefont{H.}~\bibnamefont{Hattab}}, \bibinfo {author}
  {\bibfnamefont{A.~T.}\ \bibnamefont{N'Diaye}}, \bibinfo {author}
  {\bibfnamefont{D.}~\bibnamefont{Wall}}, \bibinfo {author}
  {\bibfnamefont{C.}~\bibnamefont{Klein}}, \bibinfo {author}
  {\bibfnamefont{G.}~\bibnamefont{Jnawali}}, \bibinfo {author}
  {\bibfnamefont{J.}~\bibnamefont{Coraux}}, \bibinfo {author}
  {\bibfnamefont{C.}~\bibnamefont{Busse}}, \bibinfo {author}
  {\bibfnamefont{R.}~\bibnamefont{van Gastel}}, \bibinfo {author}
  {\bibfnamefont{B.}~\bibnamefont{Poelsema}}, \bibinfo {author}
  {\bibfnamefont{T.}~\bibnamefont{Michely}}, \bibinfo {author}
  {\bibfnamefont{F.-J.}\ \bibnamefont{Meyer~zu Heringdorf}},\ and\ \bibinfo
  {author} {\bibfnamefont{M.}~\bibnamefont{Horn-von Hoegen}},\ }%
  \bibfield{journal}{%
  \bibinfo {journal} {Nano Lett.}\ }%
  \textbf{\bibinfo {volume} {12}},\ \bibinfo {pages} {678} (\bibinfo {year}
  {2012})%
  \bibAnnoteFile{NoStop}{Hattab2012}%
\bibitem{Sampsa2013}%
  \BibitemOpen
  \bibfield{author}{%
  \bibinfo {author} {\bibfnamefont{S.~K.}\ \bibnamefont{H\"am\"al\"ainen}},
  \bibinfo {author} {\bibfnamefont{M.~P.}\ \bibnamefont{Boneschanscher}},
  \bibinfo {author} {\bibfnamefont{P.~H.}\ \bibnamefont{Jacobse}}, \bibinfo
  {author} {\bibfnamefont{I.}~\bibnamefont{Swart}}, \bibinfo {author}
  {\bibfnamefont{K.}~\bibnamefont{Pussi}}, \bibinfo {author}
  {\bibfnamefont{W.}~\bibnamefont{Moritz}}, \bibinfo {author}
  {\bibfnamefont{J.}~\bibnamefont{Lahtinen}}, \bibinfo {author}
  {\bibfnamefont{P.}~\bibnamefont{Liljeroth}},\ and\ \bibinfo {author}
  {\bibfnamefont{J.}~\bibnamefont{Sainio}},\ }%
  \bibfield{journal}{%
  \bibinfo {journal} {Phys. Rev. B}\ }%
  \textbf{\bibinfo {volume} {88}},\ \bibinfo {pages} {201406} (\bibinfo {year}
  {2013})%
  \bibAnnoteFile{NoStop}{Sampsa2013}%
\bibitem{Varykhalov2012}%
  \BibitemOpen
  \bibfield{author}{%
  \bibinfo {author} {\bibfnamefont{A.}~\bibnamefont{Varykhalov}}, \bibinfo
  {author} {\bibfnamefont{D.}~\bibnamefont{Marchenko}}, \bibinfo {author}
  {\bibfnamefont{M.~R.}\ \bibnamefont{Scholz}}, \bibinfo {author}
  {\bibfnamefont{E.~D.~L.}\ \bibnamefont{Rienks}}, \bibinfo {author}
  {\bibfnamefont{T.~K.}\ \bibnamefont{Kim}}, \bibinfo {author}
  {\bibfnamefont{G.}~\bibnamefont{Bihlmayer}}, \bibinfo {author}
  {\bibfnamefont{J.}~\bibnamefont{S\'anchez-Barriga}},\ and\ \bibinfo {author}
  {\bibfnamefont{O.}~\bibnamefont{Rader}},\ }%
  \bibfield{journal}{%
  \bibinfo {journal} {Phys. Rev. Lett.}\ }%
  \textbf{\bibinfo {volume} {108}},\ \bibinfo {pages} {066804} (\bibinfo {year}
  {2012})%
  \bibAnnoteFile{NoStop}{Varykhalov2012}%
\bibitem{Altenburg2012}%
  \BibitemOpen
  \bibfield{author}{%
  \bibinfo {author} {\bibfnamefont{S.~J.}\ \bibnamefont{Altenburg}}, \bibinfo
  {author} {\bibfnamefont{J.}~\bibnamefont{Kr\"oger}}, \bibinfo {author}
  {\bibfnamefont{T.~O.}\ \bibnamefont{Wehling}}, \bibinfo {author}
  {\bibfnamefont{B.}~\bibnamefont{Sachs}}, \bibinfo {author}
  {\bibfnamefont{A.~I.}\ \bibnamefont{Lichtenstein}},\ and\ \bibinfo {author}
  {\bibfnamefont{R.}~\bibnamefont{Berndt}},\ }%
  \bibfield{journal}{%
  \bibinfo {journal} {Phys. Rev. Lett.}\ }%
  \textbf{\bibinfo {volume} {108}},\ \bibinfo {pages} {206805} (\bibinfo {year}
  {2012})%
  \bibAnnoteFile{NoStop}{Altenburg2012}%
\bibitem{Louie1978}%
  \BibitemOpen
  \bibfield{author}{%
  \bibinfo {author} {\bibfnamefont{S.~G.}\ \bibnamefont{Louie}},\ }%
  \bibfield{journal}{%
  \bibinfo {journal} {Phys. Rev. Lett.}\ }%
  \textbf{\bibinfo {volume} {40}},\ \bibinfo {pages} {1525} (\bibinfo {year}
  {1978})%
  \bibAnnoteFile{NoStop}{Louie1978}%
\bibitem{Souk1985}%
  \BibitemOpen
  \bibfield{author}{%
  \bibinfo {author} {\bibfnamefont{P.}~\bibnamefont{Soukiassian}}, \bibinfo
  {author} {\bibfnamefont{R.}~\bibnamefont{Riwan}}, \bibinfo {author}
  {\bibfnamefont{J.}~\bibnamefont{Lecante}}, \bibinfo {author}
  {\bibfnamefont{E.}~\bibnamefont{Wimmer}}, \bibinfo {author}
  {\bibfnamefont{S.~R.}\ \bibnamefont{Chubb}},\ and\ \bibinfo {author}
  {\bibfnamefont{A.~J.}\ \bibnamefont{Freeman}},\ }%
  \bibfield{journal}{%
  \bibinfo {journal} {Phys. Rev. B}\ }%
  \textbf{\bibinfo {volume} {31}},\ \bibinfo {pages} {4911} (\bibinfo {year}
  {1985})%
  \bibAnnoteFile{NoStop}{Souk1985}%
\bibitem{Park2000}%
  \BibitemOpen
  \bibfield{author}{%
  \bibinfo {author} {\bibfnamefont{J.~Y.}\ \bibnamefont{Park}}, \bibinfo
  {author} {\bibfnamefont{U.~D.}\ \bibnamefont{Ham}}, \bibinfo {author}
  {\bibfnamefont{S.~J.}\ \bibnamefont{Kahng}}, \bibinfo {author}
  {\bibfnamefont{Y.}~\bibnamefont{Kuk}}, \bibinfo {author}
  {\bibfnamefont{K.}~\bibnamefont{Miyake}}, \bibinfo {author}
  {\bibfnamefont{K.}~\bibnamefont{Hata}},\ and\ \bibinfo {author}
  {\bibfnamefont{H.}~\bibnamefont{Shigekawa}},\ }%
  \bibfield{journal}{%
  \bibinfo {journal} {Phys. Rev. B}\ }%
  \textbf{\bibinfo {volume} {62}},\ \bibinfo {pages} {16341} (\bibinfo {year}
  {2000})%
  \bibAnnoteFile{NoStop}{Park2000}%
\bibitem{Hoevel2001}%
  \BibitemOpen
  \bibfield{author}{%
  \bibinfo {author} {\bibfnamefont{H.}~\bibnamefont{H\"ovel}}, \bibinfo
  {author} {\bibfnamefont{B.}~\bibnamefont{Grimm}},\ and\ \bibinfo {author}
  {\bibfnamefont{B.}~\bibnamefont{Reihl}},\ }%
  \bibfield{journal}{%
  \bibinfo {journal} {Surf. Sci.}\ }%
  \textbf{\bibinfo {volume} {477}},\ \bibinfo {pages} {43} (\bibinfo {year}
  {2001})%
  \bibAnnoteFile{NoStop}{Hoevel2001}%
\bibitem{Forster2003}%
  \BibitemOpen
  \bibfield{author}{%
  \bibinfo {author} {\bibfnamefont{F.}~\bibnamefont{Forster}}, \bibinfo
  {author} {\bibfnamefont{G.}~\bibnamefont{Nicolay}}, \bibinfo {author}
  {\bibfnamefont{F.}~\bibnamefont{Reinert}}, \bibinfo {author}
  {\bibfnamefont{D.}~\bibnamefont{Ehm}}, \bibinfo {author}
  {\bibfnamefont{S.}~\bibnamefont{Schmidt}},\ and\ \bibinfo {author}
  {\bibfnamefont{S.}~\bibnamefont{H\"ufner}},\ }%
  \bibfield{journal}{%
  \bibinfo {journal} {Surf. Sci.}\ }%
  \textbf{\bibinfo {volume} {532}},\ \bibinfo {pages} {160} (\bibinfo {year}
  {2003})%
  \bibAnnoteFile{NoStop}{Forster2003}%
\bibitem{Jonker1981}%
  \BibitemOpen
  \bibfield{author}{%
  \bibinfo {author} {\bibfnamefont{B.~T.}\ \bibnamefont{Jonker}}, \bibinfo
  {author} {\bibfnamefont{J.~F.}\ \bibnamefont{Morar}},\ and\ \bibinfo {author}
  {\bibfnamefont{R.~L.}\ \bibnamefont{Park}},\ }%
  \bibfield{journal}{%
  \bibinfo {journal} {Phys. Rev. B}\ }%
  \textbf{\bibinfo {volume} {24}},\ \bibinfo {pages} {2951} (\bibinfo {year}
  {1981})%
  \bibAnnoteFile{NoStop}{Jonker1981}%
\bibitem{Nicoara2006}%
  \BibitemOpen
  \bibfield{author}{%
  \bibinfo {author} {\bibfnamefont{N.}~\bibnamefont{Nicoara}}, \bibinfo
  {author} {\bibfnamefont{E.}~\bibnamefont{Roman}}, \bibinfo {author}
  {\bibfnamefont{J.~M.}\ \bibnamefont{Gomez-Rodriguez}}, \bibinfo {author}
  {\bibfnamefont{J.~A.}\ \bibnamefont{Martin-Gago}},\ and\ \bibinfo {author}
  {\bibfnamefont{J.}~\bibnamefont{Mendez}},\ }%
  \bibfield{journal}{%
  \bibinfo {journal} {Organic Electron.}\ }%
  \textbf{\bibinfo {volume} {7}},\ \bibinfo {pages} {287} (\bibinfo {year}
  {2006})%
  \bibAnnoteFile{NoStop}{Nicoara2006}%
\bibitem{Eberhardt1983}%
  \BibitemOpen
  \bibfield{author}{%
  \bibinfo {author} {\bibfnamefont{W.}~\bibnamefont{Eberhardt}}, \bibinfo
  {author} {\bibfnamefont{S.~G.}\ \bibnamefont{Louie}},\ and\ \bibinfo {author}
  {\bibfnamefont{E.~W.}\ \bibnamefont{Plummer}},\ }%
  \bibfield{journal}{%
  \bibinfo {journal} {Phys. Rev. B}\ }%
  \textbf{\bibinfo {volume} {28}},\ \bibinfo {pages} {465} (\bibinfo {year}
  {1983})%
  \bibAnnoteFile{NoStop}{Eberhardt1983}%
\bibitem{Tzeng2000}%
  \BibitemOpen
  \bibfield{author}{%
  \bibinfo {author} {\bibfnamefont{C.~T.}\ \bibnamefont{Tzeng}}, \bibinfo
  {author} {\bibfnamefont{W.~S.}\ \bibnamefont{Lo}}, \bibinfo {author}
  {\bibfnamefont{J.~Y.}\ \bibnamefont{Yuh}}, \bibinfo {author}
  {\bibfnamefont{R.~Y.}\ \bibnamefont{Chu}},\ and\ \bibinfo {author}
  {\bibfnamefont{K.~D.}\ \bibnamefont{Tsuei}},\ }%
  \bibfield{journal}{%
  \bibinfo {journal} {Phys. Rev. B}\ }%
  \textbf{\bibinfo {volume} {61}},\ \bibinfo {pages} {2263} (\bibinfo {year}
  {2000})%
  \bibAnnoteFile{NoStop}{Tzeng2000}%
\bibitem{Torrente2008}%
  \BibitemOpen
  \bibfield{author}{%
  \bibinfo {author} {\bibfnamefont{N.}~\bibnamefont{Gonzalez-Lakunza}},
  \bibinfo {author} {\bibfnamefont{I.}~\bibnamefont{Fernandez-Torrente}},
  \bibinfo {author} {\bibfnamefont{K.~J.}\ \bibnamefont{Franke}}, \bibinfo
  {author} {\bibfnamefont{N.}~\bibnamefont{Lorente}}, \bibinfo {author}
  {\bibfnamefont{A.}~\bibnamefont{Arnau}},\ and\ \bibinfo {author}
  {\bibfnamefont{J.~I.}\ \bibnamefont{Pascual}},\ }%
  \bibfield{journal}{%
  \bibinfo {journal} {Phys. Rev. Lett.}\ }%
  \textbf{\bibinfo {volume} {100}},\ \bibinfo {pages} {156805} (\bibinfo {year}
  {2008})%
  \bibAnnoteFile{NoStop}{Torrente2008}%
\bibitem{Kliewer2000}%
  \BibitemOpen
  \bibfield{author}{%
  \bibinfo {author} {\bibfnamefont{J.}~\bibnamefont{Kliewer}}, \bibinfo
  {author} {\bibfnamefont{R.}~\bibnamefont{Berndt}}, \bibinfo {author}
  {\bibfnamefont{E.~V.}\ \bibnamefont{Chulkov}}, \bibinfo {author}
  {\bibfnamefont{V.~M.}\ \bibnamefont{Silkin}}, \bibinfo {author}
  {\bibfnamefont{P.~M.}\ \bibnamefont{Echenique}},\ and\ \bibinfo {author}
  {\bibfnamefont{S.}~\bibnamefont{Crampin}},\ }%
  \bibfield{journal}{%
  \bibinfo {journal} {Science}\ }%
  \textbf{\bibinfo {volume} {288}},\ \bibinfo {pages} {1399} (\bibinfo {year}
  {2000})%
  \bibAnnoteFile{NoStop}{Kliewer2000}%
\bibitem{Binnig1985}%
  \BibitemOpen
  \bibfield{author}{%
  \bibinfo {author} {\bibfnamefont{G.}~\bibnamefont{Binnig}}, \bibinfo {author}
  {\bibfnamefont{K.~H.}\ \bibnamefont{Frank}}, \bibinfo {author}
  {\bibfnamefont{H.}~\bibnamefont{Fuchs}}, \bibinfo {author}
  {\bibfnamefont{N.}~\bibnamefont{Garcia}}, \bibinfo {author}
  {\bibfnamefont{B.}~\bibnamefont{Reihl}}, \bibinfo {author}
  {\bibfnamefont{H.}~\bibnamefont{Rohrer}}, \bibinfo {author}
  {\bibfnamefont{F.}~\bibnamefont{Salvan}},\ and\ \bibinfo {author}
  {\bibfnamefont{A.~R.}\ \bibnamefont{Williams}},\ }%
  \bibfield{journal}{%
  \bibinfo {journal} {Phys. Rev. Lett.}\ }%
  \textbf{\bibinfo {volume} {55}},\ \bibinfo {pages} {991} (\bibinfo {year}
  {1985})%
  \bibAnnoteFile{NoStop}{Binnig1985}%
\bibitem{Becker1985}%
  \BibitemOpen
  \bibfield{author}{%
  \bibinfo {author} {\bibfnamefont{R.~S.}\ \bibnamefont{Becker}}, \bibinfo
  {author} {\bibfnamefont{J.~A.}\ \bibnamefont{Golovchenko}},\ and\ \bibinfo
  {author} {\bibfnamefont{B.~S.}\ \bibnamefont{Swartzentruber}},\ }%
  \bibfield{journal}{%
  \bibinfo {journal} {Phys. Rev. Lett.}\ }%
  \textbf{\bibinfo {volume} {55}},\ \bibinfo {pages} {987} (\bibinfo {year}
  {1985})%
  \bibAnnoteFile{NoStop}{Becker1985}%
\bibitem{Gundlach1966}%
  \BibitemOpen
  \bibfield{author}{%
  \bibinfo {author} {\bibfnamefont{K.~H.}\ \bibnamefont{Gundlach}},\ }%
  \bibfield{journal}{%
  \bibinfo {journal} {Solid State Electron.}\ }%
  \textbf{\bibinfo {volume} {9}},\ \bibinfo {pages} {949} (\bibinfo {year}
  {1966})%
  \bibAnnoteFile{NoStop}{Gundlach1966}%
\bibitem{Fowler1928}%
  \BibitemOpen
  \bibfield{author}{%
  \bibinfo {author} {\bibfnamefont{R.~H.}\ \bibnamefont{Fowler}}\ and\ \bibinfo
  {author} {\bibfnamefont{L.}~\bibnamefont{Nordheim}},\ }%
  \bibfield{journal}{%
  \bibinfo {journal} {Proc. R. Soc. A}\ }%
  \textbf{\bibinfo {volume} {119}},\ \bibinfo {pages} {173} (\bibinfo {year}
  {1928})%
  \bibAnnoteFile{NoStop}{Fowler1928}%
\bibitem{Koles2000}%
  \BibitemOpen
  \bibfield{author}{%
  \bibinfo {author} {\bibfnamefont{O.~Y.}\ \bibnamefont{Kolesnychenko}},
  \bibinfo {author} {\bibfnamefont{Y.~A.}\ \bibnamefont{Kolesnichenko}},
  \bibinfo {author} {\bibfnamefont{O.}~\bibnamefont{Shklyarevskii}},\ and\
  \bibinfo {author} {\bibfnamefont{H.}~\bibnamefont{van Kempen}},\ }%
  \bibfield{journal}{%
  \bibinfo {journal} {Physica B}\ }%
  \textbf{\bibinfo {volume} {291}},\ \bibinfo {pages} {246} (\bibinfo {year}
  {2000})%
  \bibAnnoteFile{NoStop}{Koles2000}%
\bibitem{Rienks2005}%
  \BibitemOpen
  \bibfield{author}{%
  \bibinfo {author} {\bibfnamefont{E.~D.~L.}\ \bibnamefont{Rienks}}, \bibinfo
  {author} {\bibfnamefont{N.}~\bibnamefont{Nilius}}, \bibinfo {author}
  {\bibfnamefont{H.~P.}\ \bibnamefont{Rust}},\ and\ \bibinfo {author}
  {\bibfnamefont{H.~J.}\ \bibnamefont{Freund}},\ }%
  \bibfield{journal}{%
  \bibinfo {journal} {Phys. Rev. B}\ }%
  \textbf{\bibinfo {volume} {71}},\ \bibinfo {pages} {241404} (\bibinfo {year}
  {2005})%
  \bibAnnoteFile{NoStop}{Rienks2005}%
\bibitem{Koenig2009}%
  \BibitemOpen
  \bibfield{author}{%
  \bibinfo {author} {\bibfnamefont{T.}~\bibnamefont{K\"onig}}, \bibinfo
  {author} {\bibfnamefont{G.~H.}\ \bibnamefont{Simon}}, \bibinfo {author}
  {\bibfnamefont{H.~P.}\ \bibnamefont{Rust}},\ and\ \bibinfo {author}
  {\bibfnamefont{M.}~\bibnamefont{Heyde}},\ }%
  \bibfield{journal}{%
  \bibinfo {journal} {J. Phys. Chem. C}\ }%
  \textbf{\bibinfo {volume} {113}},\ \bibinfo {pages} {11301} (\bibinfo {year}
  {2009})%
  \bibAnnoteFile{NoStop}{Koenig2009}%
\bibitem{Pivetta2005}%
  \BibitemOpen
  \bibfield{author}{%
  \bibinfo {author} {\bibfnamefont{M.}~\bibnamefont{Pivetta}}, \bibinfo
  {author} {\bibfnamefont{F.}~\bibnamefont{Patthey}}, \bibinfo {author}
  {\bibfnamefont{M.}~\bibnamefont{Stengel}}, \bibinfo {author}
  {\bibfnamefont{A.}~\bibnamefont{Baldereschi}},\ and\ \bibinfo {author}
  {\bibfnamefont{W.~D.}\ \bibnamefont{Schneider}},\ }%
  \bibfield{journal}{%
  \bibinfo {journal} {Phys. Rev. B}\ }%
  \textbf{\bibinfo {volume} {72}},\ \bibinfo {pages} {115404} (\bibinfo {year}
  {2005})%
  \bibAnnoteFile{NoStop}{Pivetta2005}%
\bibitem{Ploigt2007}%
  \BibitemOpen
  \bibfield{author}{%
  \bibinfo {author} {\bibfnamefont{H.-C.}\ \bibnamefont{Ploigt}}, \bibinfo
  {author} {\bibfnamefont{C.}~\bibnamefont{Brun}}, \bibinfo {author}
  {\bibfnamefont{M.}~\bibnamefont{Pivetta}}, \bibinfo {author}
  {\bibfnamefont{F.}~\bibnamefont{Patthey}},\ and\ \bibinfo {author}
  {\bibfnamefont{W.-D.}\ \bibnamefont{Schneider}},\ }%
  \bibfield{journal}{%
  \bibinfo {journal} {Phys. Rev. B}\ }%
  \textbf{\bibinfo {volume} {76}},\ \bibinfo {pages} {195404} (\bibinfo {year}
  {2007})%
  \bibAnnoteFile{NoStop}{Ploigt2007}%
\bibitem{Ruggiero2007}%
  \BibitemOpen
  \bibfield{author}{%
  \bibinfo {author} {\bibfnamefont{C.~D.}\ \bibnamefont{Ruggiero}}, \bibinfo
  {author} {\bibfnamefont{T.}~\bibnamefont{Choi}},\ and\ \bibinfo {author}
  {\bibfnamefont{J.~A.}\ \bibnamefont{Gupta}},\ }%
  \bibfield{journal}{%
  \bibinfo {journal} {Appl. Phys. Lett.}\ }%
  \textbf{\bibinfo {volume} {91}},\ \bibinfo {pages} {253106} (\bibinfo {year}
  {2007})%
  \bibAnnoteFile{NoStop}{Ruggiero2007}%
\bibitem{Wang2010}%
  \BibitemOpen
  \bibfield{author}{%
  \bibinfo {author} {\bibfnamefont{B.}~\bibnamefont{Wang}}, \bibinfo {author}
  {\bibfnamefont{M.}~\bibnamefont{Caffio}}, \bibinfo {author}
  {\bibfnamefont{C.}~\bibnamefont{Bromley}}, \bibinfo {author}
  {\bibfnamefont{H.}~\bibnamefont{Fruechtl}},\ and\ \bibinfo {author}
  {\bibfnamefont{R.}~\bibnamefont{Schaub}},\ }%
  \bibfield{journal}{%
  \bibinfo {journal} {ACS Nano}\ }%
  \textbf{\bibinfo {volume} {4}},\ \bibinfo {pages} {5773} (\bibinfo {year}
  {2010})%
  \bibAnnoteFile{NoStop}{Wang2010}%
\bibitem{Michaelson1977}%
  \BibitemOpen
  \bibfield{author}{%
  \bibinfo {author} {\bibfnamefont{H.~B.}\ \bibnamefont{Michaelson}},\ }%
  \bibfield{journal}{%
  \bibinfo {journal} {J. Appl. Phys.}\ }%
  \textbf{\bibinfo {volume} {48}},\ \bibinfo {pages} {4729} (\bibinfo {year}
  {1977})%
  \bibAnnoteFile{NoStop}{Michaelson1977}%
\bibitem{Lin2007}%
  \BibitemOpen
  \bibfield{author}{%
  \bibinfo {author} {\bibfnamefont{C.~L.}\ \bibnamefont{Lin}}, \bibinfo
  {author} {\bibfnamefont{S.~M.}\ \bibnamefont{Lu}}, \bibinfo {author}
  {\bibfnamefont{W.~B.}\ \bibnamefont{Su}}, \bibinfo {author}
  {\bibfnamefont{H.~T.}\ \bibnamefont{Shih}}, \bibinfo {author}
  {\bibfnamefont{B.~F.}\ \bibnamefont{Wu}}, \bibinfo {author}
  {\bibfnamefont{Y.~D.}\ \bibnamefont{Yao}}, \bibinfo {author}
  {\bibfnamefont{C.~S.}\ \bibnamefont{Chang}},\ and\ \bibinfo {author}
  {\bibfnamefont{T.~T.}\ \bibnamefont{Tsong}},\ }%
  \bibfield{journal}{%
  \bibinfo {journal} {Phys. Rev. Lett.}\ }%
  \textbf{\bibinfo {volume} {99}},\ \bibinfo {pages} {216103} (\bibinfo {year}
  {2007})%
  \bibAnnoteFile{NoStop}{Lin2007}%
\bibitem{Kern2010}%
  \BibitemOpen
  \bibfield{author}{%
  \bibinfo {author} {\bibfnamefont{L.}~\bibnamefont{Vitali}}, \bibinfo {author}
  {\bibfnamefont{G.}~\bibnamefont{Levita}}, \bibinfo {author}
  {\bibfnamefont{R.}~\bibnamefont{Ohmann}}, \bibinfo {author}
  {\bibfnamefont{A.}~\bibnamefont{Comisso}}, \bibinfo {author}
  {\bibfnamefont{A.}~\bibnamefont{De~Vita}},\ and\ \bibinfo {author}
  {\bibfnamefont{K.}~\bibnamefont{Kern}},\ }%
  \bibfield{journal}{%
  \bibinfo {journal} {Nat. Mater.}\ }%
  \textbf{\bibinfo {volume} {9}},\ \bibinfo {pages} {320} (\bibinfo {year}
  {2010})%
  \bibAnnoteFile{NoStop}{Kern2010}%
\bibitem{Ishii1999}%
  \BibitemOpen
  \bibfield{author}{%
  \bibinfo {author} {\bibfnamefont{H.}~\bibnamefont{Ishii}}, \bibinfo {author}
  {\bibfnamefont{K.}~\bibnamefont{Sugiyama}}, \bibinfo {author}
  {\bibfnamefont{E.}~\bibnamefont{Ito}},\ and\ \bibinfo {author}
  {\bibfnamefont{K.}~\bibnamefont{Seki}},\ }%
  \bibfield{journal}{%
  \bibinfo {journal} {Adv. Mater.}\ }%
  \textbf{\bibinfo {volume} {11}},\ \bibinfo {pages} {605} (\bibinfo {year}
  {1999})%
  \bibAnnoteFile{NoStop}{Ishii1999}%
\bibitem{Hwang2009}%
  \BibitemOpen
  \bibfield{author}{%
  \bibinfo {author} {\bibfnamefont{J.}~\bibnamefont{Hwang}}, \bibinfo {author}
  {\bibfnamefont{A.}~\bibnamefont{Wan}},\ and\ \bibinfo {author}
  {\bibfnamefont{A.}~\bibnamefont{Kahn}},\ }%
  \bibfield{journal}{%
  \bibinfo {journal} {Mater. Sci. Eng. R-Rep.}\ }%
  \textbf{\bibinfo {volume} {64}},\ \bibinfo {pages} {1} (\bibinfo {year}
  {2009})%
  \bibAnnoteFile{NoStop}{Hwang2009}%
\bibitem{Smoluchowski1941}%
  \BibitemOpen
  \bibfield{author}{%
  \bibinfo {author} {\bibfnamefont{R.}~\bibnamefont{Smoluchowski}},\ }%
  \bibfield{journal}{%
  \bibinfo {journal} {Phys. Rev.}\ }%
  \textbf{\bibinfo {volume} {60}},\ \bibinfo {pages} {661} (\bibinfo {year}
  {1941})%
  \bibAnnoteFile{NoStop}{Smoluchowski1941}%
\bibitem{Lang1971}%
  \BibitemOpen
  \bibfield{author}{%
  \bibinfo {author} {\bibfnamefont{N.~D.}\ \bibnamefont{Lang}}\ and\ \bibinfo
  {author} {\bibfnamefont{W.}~\bibnamefont{Kohn}},\ }%
  \bibfield{journal}{%
  \bibinfo {journal} {Phys. Rev. B}\ }%
  \textbf{\bibinfo {volume} {3}},\ \bibinfo {pages} {1215} (\bibinfo {year}
  {1971})%
  \bibAnnoteFile{NoStop}{Lang1971}%
\bibitem{Nieminen1976}%
  \BibitemOpen
  \bibfield{author}{%
  \bibinfo {author} {\bibfnamefont{R.~M.}\ \bibnamefont{Nieminen}}\ and\
  \bibinfo {author} {\bibfnamefont{C.~H.}\ \bibnamefont{Hodges}},\ }%
  \bibfield{journal}{%
  \bibinfo {journal} {J. Phys. F}\ }%
  \textbf{\bibinfo {volume} {6}},\ \bibinfo {pages} {573} (\bibinfo {year}
  {1976})%
  \bibAnnoteFile{NoStop}{Nieminen1976}%
\end{thebibliography}
\end{document}